\documentclass{aa}
\usepackage{graphics,epsf,epsfig,graphicx}

\begin{document}

\title{Formation of polar ring galaxies}
                                
\author{Fr\'ed\'eric Bournaud \inst{1,2} \and  Fran\c{c}oise Combes \inst{1}} 

\offprints{F. Bournaud, \email{Frederic.Bournaud@obspm.fr}} 
\institute{Observatoire de Paris, LERMA, 61 Av. de l'Observatoire, 
F-75014, Paris, France 
\and
Ecole Normale Sup\'erieure, 45 rue d'Ulm, F-75005, Paris, France} 
\date{Received 2 october 2002; accepted 17 january 2003}
\authorrunning{F. Bournaud \& F. Combes}
\titlerunning{Formation of polar ring galaxies}

\abstract{Polar ring galaxies are peculiar systems in which a gas-rich, nearly polar ring surrounds an early-type or elliptical host galaxy. Two formation scenarios for these objects have been proposed: they are thought to form either in major galaxy mergers or by tidal accretion of the polar material from a gas rich donor galaxy. Both scenarios are studied through N-body simulations including gas dynamics and star formation. Constraints on physical parameters are drawn out, in order to determine which scenario is the most likely to occur. Polar ring galaxies from each scenario are compared with observations and we discuss whether the accretion scenario and the merging scenario account for observational properties of polar ring galaxies. The conclusion of this study is that the accretion scenario is both the most likely and the most supported by observations. Even if the merging scenario is rather robust, most polar ring galaxies are shown to be the result of tidal gas accretion events.
}

\maketitle

\section{Introduction}

A polar ring galaxy (PRG) is made up of an early-type, lenticular, or elliptical host galaxy, surrounded by a ring of gas and stars orbiting in a plane that is nearly polar (Whitmore et al. 1990). The presence of two dynamical planes is thought to give an insight into the shape of dark halos in such systems, which explains why PRGs are interesting peculiar systems. The prototype polar ring galaxy, NGC~4650A, has been the target of several kinematical studies: Whitmore et al. (1987) concluded that the dark matter halo is nearly spherical. Sackett \& Sparke (1990) and Sackett et al. (1994) have proposed a dynamical model that indicates that the dark halo is flattened, with its major axis along the equatorial plane of the host galaxy. More recently, Combes \& Arnaboldi (1996) have built a different model in which the dark matter is only needed to explain velocities at large radii in the polar ring, while the velocities in the host galaxy are accounted for by the visible mass; thus they concluded that the dark halo of NGC~4650A may be flattened towards the polar ring. In all these studies the common conclusion was that PRG were embedded in a dark halo, for the visible mass is not enough to account for the observed velocities in rings. According to a different approach based on the Tully-Fisher relation, applied to several PRGs, Iodice et al. (2002c) have argued that the dark halo may be flattened along the polar ring plane in most PRGs.

From the shape of the dark matter distribution in PRGs, we would like to get information on dark halos around spiral galaxies in general. At least for this reason, it seems fundamental to know how polar rings form. Polar rings are usually thought to be formed during a secondary event around a pre-existing galaxy. The collapse of a protogalactic cloud could create two misaligned systems (Curir \& Diaferio 1994), thus the polar ring and the host galaxy might be formed at the same time. However, polar rings appear to be younger than host galaxies: they are gas-rich (see HI surveys from Richter et al. 1994 and van Driel et al. 2000, 2002), while the host galaxy is generally depleted of gas; moreover polar rings contain a young stellar population while host galaxies contain old stars, as indicated by their color indices (Iodice 2002b, 2002c). Thus, it seems reasonable to admit that PRGs are made up of a previously-formed host galaxy and a more recent polar structure.

Two kinds of scenarios have then been proposed to account for the formation of a polar ring around a preexisting galaxy (Iodice 2001):
\begin{itemize}
\item the merging scenario: proposed by Bekki (1997, 1998), this scenario assumes a head-on collision between two orthogonal spiral galaxies (see Fig.~\ref{schem_merging}). The first one is called the intruder, the second one the victim. When the relative velocity of colliding galaxies is large, such collisions are known to form cartwheel-like rings (Horellou \& Combes, 2001) that are not polar rings for they do not surround a host galaxy and are transient expanding features. When the relative velocity is lower, Bekki (1997, 1998) has shown that the two galaxies merge, owing to dynamical friction: the intruder becomes the host galaxy, while the gas content of the victim galaxy forms the polar ring. 
	\begin{figure}
	\centering
	\includegraphics[angle=270,width=8cm]{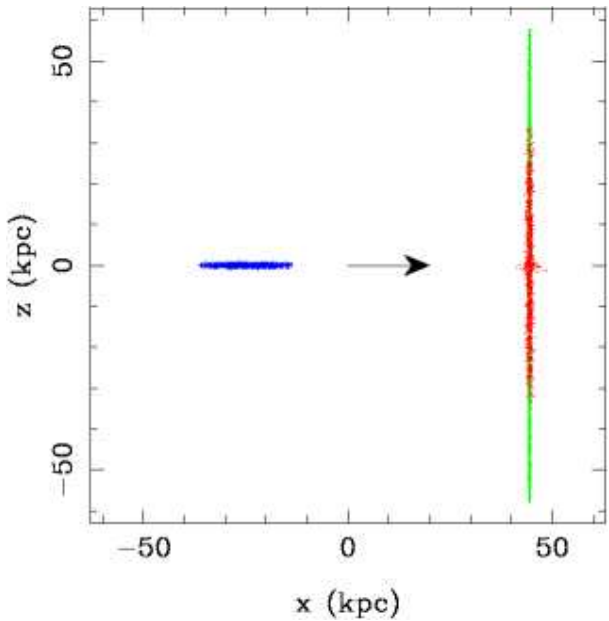}
	\caption{Initial configuration of a polar ring forming galaxy merger (Bekki, 1997). The intruder galaxy is on the left. The victim galaxy is gas rich, and contains gas at radii larger than the stellar disk radius.}
	\label{schem_merging}	
	\end{figure}
\item the accretion scenario: this scenario (e.g. Schweizer et al. 1983, Reshetnikov \& Sotnikova 1997) consists of the accretion of gas from another galaxy by the host. Our galaxy has a faint polar ring accreted from the large Magellanic cloud (Haud, 1988). Rings around other galaxies are more massive, and would rather be accreted from massive gas-rich galaxies, i.e. late-type spirals. This scenario assumes that two galaxies interact, yet no merging is needed: tidal interactions can account for gas accretion without any merging event. The two interacting galaxies may also merge after the formation of the ring. The formation of a polar ring requires that the donor galaxy is on a nearly polar orbit with respect to the accreting host galaxy.
\end{itemize}
It has been demonstrated that both scenarios can explain the formation of polar rings surrounding preexisting galaxies (Bekki 1997, Reshetnikov \& Sotnikova 1997). Yet, they have not been compared in the same work, and the stability of such rings, probably related to their self gravity (Sparke 1986, Arnaboldi \& Sparke 1994), still has to be addressed with regard to the formation scenario (Iodice et al., 2002a).

In this paper, we study both scenarios through the same numerical model. We wish to confirm that they both form stable polar rings, and to determine which among them is the most robust. We will then address the dependence of both scenarios on physical parameters, in order to compare the probability of ring formation according to each one. We will also compare the rings that each kind of event creates with observations of PRGs.

PRGs were first classified by Whitmore et al. (1990). From these observations, and more recent ones (van Driel et al. 2000, Iodice et al. 2002a, 2002b, Arnaboldi et al. 1993, van Driel et al. 1995), we can summarize the main properties of PRGs that a formation scenario has to account for (Iodice 2001):
\begin{itemize}
\item Host galaxies are often SO or early-type galaxies: they morphologically look like SOs, but their photometry is typical of early-type spirals, as revealed by Iodice et al. (2002b, 2002c). Some are ellipticals, as in AM~2020-504, or late-type gas rich spirals, as in NGC~660.
\item About 0.5\% of the possible host galaxies in the local Universe (SO and early-type) are observed to have a polar ring or a robust polar ring candidate. Yet, only a specific line-of-sight makes it possible to detect the ring, so that Whitmore et al. (1990) estimate the actual percentage to about 4.5\%. It could be even greater if one also accounted for polar rings that may have been destroyed by more recent galaxy interactions. This percentage does not consider the faintest polar rings, such as the ring of our own galaxy, so that 4.5\% is a lower limit to the actual percentage.
\item Polar rings may be about as massive as the host galaxy, as well as much lighter.
\item Radii of polar rings can be much larger than the host scale-length, as well as of the same order.
\item Most rings are inclined by less than 25 degrees from the polar axis of the host galaxy, but some highly inclined rings are observed, for instance NGC~660, in which the ring inclination is about 45 degrees (see van Driel et al. 1995).
\item Some polar rings contain an old stellar population, which indicates that they are stable structures with a lifetime of at least a few Gyrs.
\item Polar rings show various inner morphologies, such as helicoidal rings, double rings, or rings with spiral arms.
\end{itemize}

In Sect.~\ref{s2} we describe the numerical schemes that are used in this work. In Sect.~\ref{s3} and \ref{s4} we study the merging scenario and the accretion one respectively. We compare them in Sect.~\ref{s5}, \ref{s6}, and \ref{s7}. We conclude in Sect.~\ref{s8}.

\section{Numerical techniques}\label{s2}

In this section, we describe the numerical schemes and the mass models for spiral and elliptical galaxies that we will use to study formation scenarios. The N-body particle code includes gas dynamics, star formation and stellar mass-loss.

Stars, gas and dark matter are described by 2 to 20$\cdot$10$^5$ particles. The gravitational potential is deduced from the particle positions via a FFT method. The particles are first meshed on a Cartesian grid (generally $128\times128\times128$ cells), according to the multilinear ``cloud-in-cell'' (CIC) interpolation. The convolution of the density and the Green function is computed via FFT techniques (James, 1977). Two-body relaxation is overestimated in particle codes because of the small number of particles. To suppress the two-body relaxation, the gravitational forces are softened with a characteristic size of 300 to 1000pc, depending on the grid size.

The dissipative nature of the interstellar medium is modeled with a sticky-particles algorithm that reproduces inelastic cloud-cloud collisions (Schwarz, 1981). The relative kinetic energy of gas clouds is lost, and the gas re-radiates this energy away (Combes \& Gerin 1985). Collisions are detected on a grid, and when two particles collide, their radial (along the line joining them) and tangential relative velocities are reduced by factors $\beta_r$ and $\beta_t$. We generally choose $\beta_r=0.65$ and $\beta_t=0.65$. A detailed description of our sticky-particles code and a discussion of $\beta_r$ and $\beta_t$ is given in Bournaud \& Combes (2002).

The code also includes numerical schemes for star formation and stellar mass-loss that are described in Bournaud \& Combes (2002).

The models for spiral and elliptical galaxies and dark halos are described in Appendix~\ref{append}.
\section{Merging scenario}\label{s3}

\subsection{Numerical simulations and relevant parameters}
There are many free parameters for the merging scenario: the parameters for each disk, bulge, halo, and the orbital parameters. Some numerical tests have allowed us to determine which parameters may influence the formation of polar rings and the properties of polar rings after their formation. For instance, the bulge-to-disk mass ratio is not a relevant parameter, and we will not vary it. The parameters that have a significant influence are:
\begin{itemize}
	\item the stellar and gas mass in both galaxies (intruder and victim).
	\item the radius of the stellar disk and the gas radial extent in both galaxies.
	\item the dark halo shape.
	\item the relative velocities of galaxies before the collision, $V_{coll}$.
	\item the angle between the two disk planes, $\Theta$.
	\item the angle between the orbital plane of the intruder and the victim disk, $\Phi$.
	\item the radius at which the intruder crosses the victim disk, $R_{coll}$, in units of the victim disk radius.
\end{itemize}

\begin{figure}
	\centering
	\includegraphics[angle=0,width=8cm]{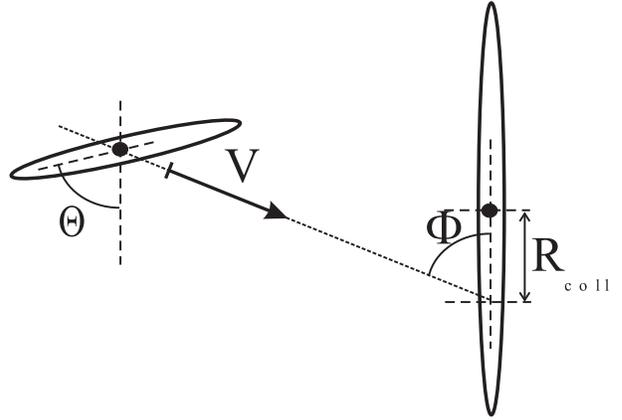}
	\caption{Collisional parameters for the merging scenario. The intruder is on the left, the victim is on the right. Ideal values of these parameters are $\Theta=90$, $\Phi=90$, $R_{coll}=0$. $R_{coll}$ is always given in units of the disk radius of the victim.}\label{param_merging}
	\end{figure}

Parameters $\Theta$, $\Phi$, and $R_{coll}$ are illustrated in Fig.~\ref{param_merging}.

	\begin{table*}
	\centering
	\begin{minipage}{18cm}
	\centering
	\begin{tabular}{ccccccccccccc}
	\hline
	\hline
	Run & M$^S_{\mathrm{intr}}$ & M$^G_{\mathrm{intr}}$ &
	R$^G_{\mathrm{intr}}$ & M$^S_{\mathrm{vict}}$ & M$^G_{\mathrm{vict}}$ &
	R$^G_{\mathrm{vict}}$ & E$_{\mathrm{intr}}$ & E$_{\mathrm{vict}}$ &
	V$_{coll}$ & $\Theta$ & $\Phi$ & $R_{coll}$ \\
	&
	\footnote{Stellar mass (intruder)}&
	\footnote{Gas mass (intruder)}&
	\footnote{External radius of the gaseous disk in units of stellar radius (intruder)}&
	\footnote{Stellar mass (victim)}&
	\footnote{Gas mass (victim)}&
	\footnote{External radius of the gaseous disk in units of stellar radius (victim)}&
	\footnote{Dark halo shape (intruder); E0: spherical; E$x$: spheroidal with ellipticity 1-$x/10$.}&
	\footnote{Dark halo shape (victim)}&
	\footnote{Relative velocity of the colliding galaxies}&
	\footnote{Angle between the two disks}&
	\footnote{Angle of the intruder's path on the victim's plane}&
	\footnote{Radius at which the intruder crosses the victim disk}\\
	\hline

A1  & 10 & 0 & - & 10 &3.3& 1   & E0 & E0 & 30  & 90 & 90 & 0   \\
A2  & 7  & 0 & - & 14 &3.5& 1   & E3 & E3 & 30  & 90 & 90 & 0   \\
A3  & 7  & 0 & - & 14 &3.5& 1   & E3 & E3 & 45  & 90 & 90 & 0   \\
A4  & 6  & 0 & - & 18 & 5 & 1   & E5 & E5 & 40  & 90 & 90 & 0   \\
A5  & 12 & 0 & - & 14 &3.5& 1   & E4 & E4 & 40  & 90 & 90 & 0   \\
A6  & 12 & 0 & - & 14 &3.5& 1   & E4 & E4 & 80  & 90 & 90 & 0   \\
A7  & 12 & 0 & - & 14 & 5 & 1   & E9 & E9 & 80  & 90 & 90 & 0   \\
A8  & 12 & 0 & - & 14 & 5 & 1   & E0 & E0 & 120 & 90 & 90 & 0   \\
A9  & 20 & 0 & - & 18 & 6 & 3   & E3 & E3 & 120 & 90 & 90 & 0   \\
A10 & 15 & 0 & - & 14 & 6 & 7   & E3 & E3 & 120 & 90 & 90 & 0   \\
A11 & 15 & 0 & - & 10 & 6 & 5.5 & E3 & E3 & 120 & 90 & 90 & 0   \\
A12 & 15 & 0 & - & 15 & 10& 2.5 & E3 & E3 & 120 & 90 & 90 & 0   \\
A13 & 15 & 0 & - & 15 & 10& 2.5 & E9 & E9 & 120 & 90 & 90 & 0   \\
B1  & 10 & 0 & - & 12 & 6 & 3   & E3 & E3 & 65  & 70 & 90 & 0   \\
B2  & 10 & 0 & - & 12 & 6 & 3   & E3 & E3 & 65  & 55 & 90 & 0   \\
B3  & 10 & 0 & - & 12 & 6 & 3   & E3 & E3 & 65  & 45 & 90 & 0   \\
B4  & 10 & 0 & - & 12 & 6 & 3   & E3 & E3 & 65  & 90 & 60 & 0  \\
B5  & 10 & 0 & - & 12 & 6 & 3   & E3 & E3 & 65  & 90 & 40 & 0  \\
B6  & 10 & 0 & - & 12 & 6 & 3   & E3 & E3 & 65  & 90 & 90 & .3 \\
B7  & 10 & 0 & - & 12 & 6 & 3   & E3 & E3 & 65  & 90 & 90 & .6 \\
B8  & 20 & 0 & - & 15 & 7 & 3.5 & E5 & E5 & 120 & 90 & 90 & 0   \\
B9  & 20 & 0 & - & 15 & 7 & 3.5 & E5 & E5 & 120 & 70 & 90 & 0   \\
B10 & 20 & 0 & - & 15 & 7 & 3.5 & E5 & E5 & 120 & 60 & 90 & 0   \\
B11 & 20 & 0 & - & 15 & 7 & 3.5 & E5 & E5 & 120 & 50 & 90 & 0   \\
B12 & 15 & 0 & - & 15 & 6 & 3   & E5 & E5 & 100 & 90 & 90 & 0   \\ 
B13 & 15 & 2 & 1 & 15 & 6 & 4   & E5 & E5 & 100 & 90 & 90 & 0   \\
B14 & 15 & 2 & 2 & 15 & 6 & 4   & E5 & E5 & 100 & 90 & 90 & 0   \\
B15 & 15 & 4 & 1 & 15 & 6 & 4   & E5 & E5 & 100 & 90 & 90 & 0   \\
B16 & 15 & 4 & 2 & 15 & 6 & 4   & E5 & E5 & 100 & 90 & 90 & 0   \\
B17 & 15 & 0 & - & 15 & 6 & 2   & E5 & E5 & 100 & 90 & 90 & .3  \\
B18 & 15 & 0 & - & 15 & 6 & 2   & E5 & E5 & 100 & 90 & 90 & .5  \\
B19 & 15 & 0 & - & 15 & 6 & 2   & E5 & E5 & 100 & 90 & 90 & .7  \\
B20 & 15 & 0 & - & 15 & 6 & 3   & E5 & E5 & 100 & 90 & 45 & 0  \\
B21 & 15 & 0 & - & 15 & 6 & 3   & E5 & E5 & 100 & 90 & 35 & 0  \\

	\hline
	\end{tabular}
	\caption{Run parameters. Units: $10^9$ M$_{\sun}$, kpc, km.s$^{-1}$}\label{par}
	\end{minipage}
	\end{table*}

We give in Table~\ref{par} the values of the most relevant parameters for each numerical run. Runs A1 to A13 aim at studying the influence of the structure of each galaxy and the morphology of the PRG. Runs B1 to B21 study the influence of collisional parameters in order to determine the probability that a major galaxy merger results in a polar ring.

\subsection{General properties}
	\subsubsection{Morphology of the polar ring}
Polar rings are obtained in runs A1 to A13. A double ring is obtained in run A2. Spiral arms are observed in runs A5 and A11. In run A4 we obtain a polar structure that may be regarded as a polar disk: the center of the polar structure is partly but not totally depleted. In Table~\ref{resA} we give the radius of each ring, defined by:
 \begin{equation}\label{eqR}
 R=\frac{\int_0^\infty r^2 \mu(r) dr}{\int_0^\infty r \mu(r) dr}
 \end{equation}
where $\mu(r)$ is the mean surface density of the polar ring. When we compute the radius of polar rings, we only account for the matter of the polar ring, i.e. the matter that was initially the gas of the victim. The intruder's components and the victim's stars are ignored.

           \begin{table}
	   \centering
           \begin{tabular}{cc}
           \hline
 	   \hline
           Run & ring radius $R$ \\
           \hline
           A1 & 2.5 kpc  \\
           A2 & 2.5 kpc  \\
           A3 & 3 kpc  \\
           A4 & 6 kpc  \\
           A5 & 4 kpc  \\
           A6 & 3.5 kpc   \\
           A7 & 3.5 kpc  \\ 
           A8 & 3.5 kpc  \\          
           A9 & 15 kpc  \\
           A10 & 35 kpc  \\
           A11 & 27 kpc \\
           A12 & 13 kpc  \\
           A13 & 14 kpc  \\
	   \hline
           \end{tabular}
           \caption{Ring radius in runs A1 to A13.}\label{resA}
           \end{table}

\begin{figure*}
	\centering
	\includegraphics[angle=270,width=6cm]{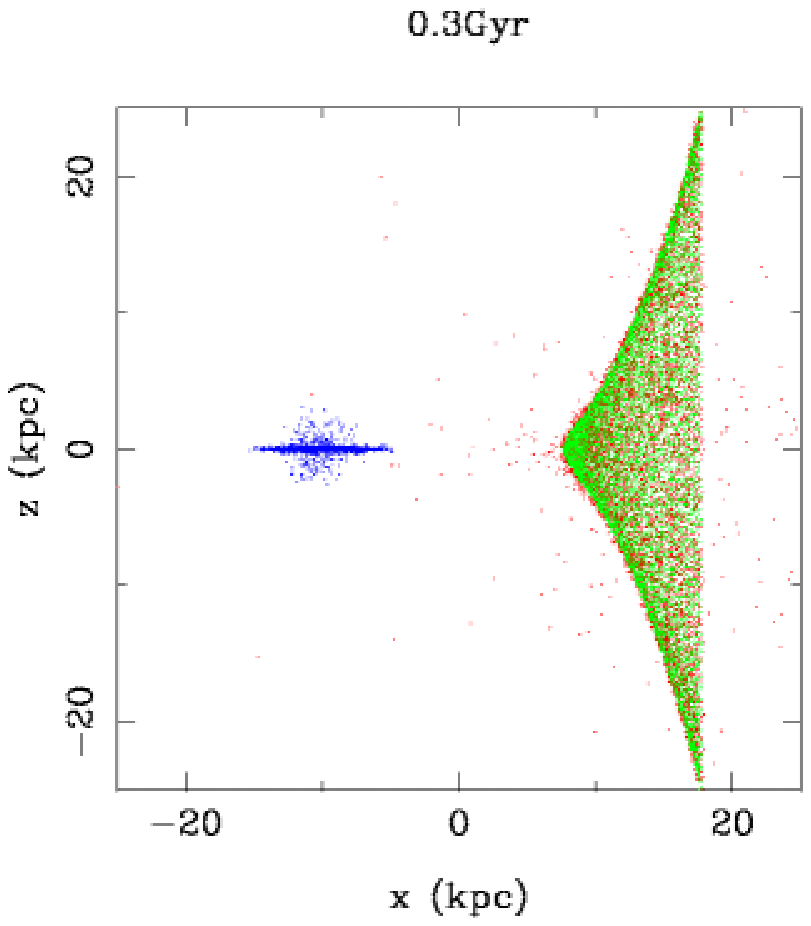}
	\hspace{.3cm}
	\includegraphics[angle=270,width=6cm]{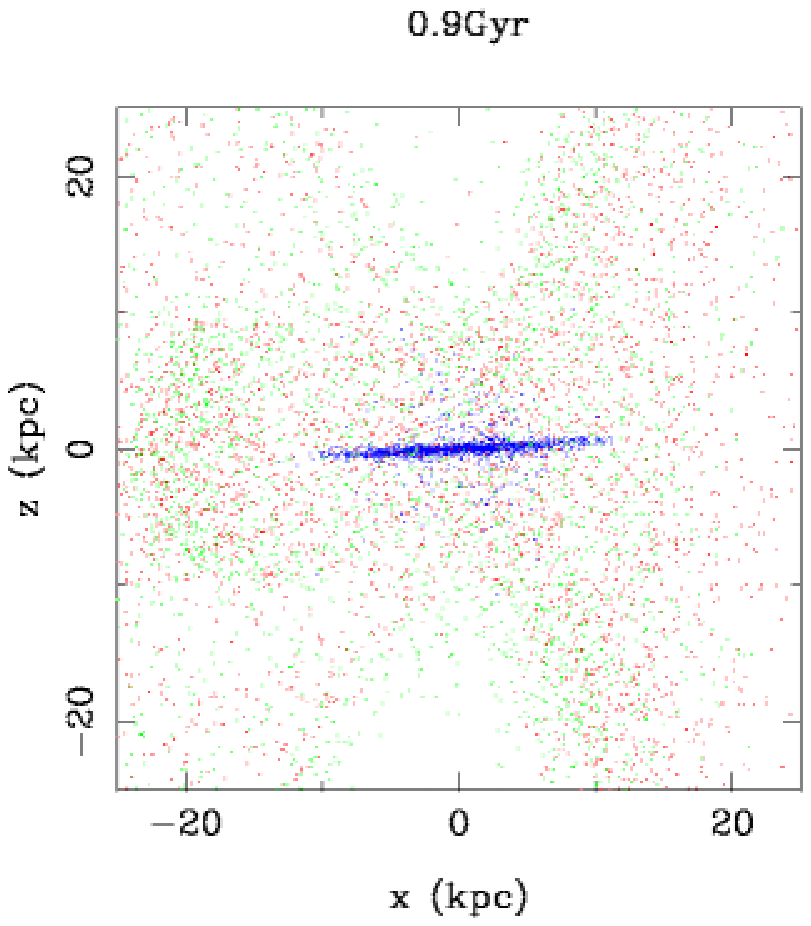}\\
	\vspace{.3cm}
	\includegraphics[angle=270,width=6cm]{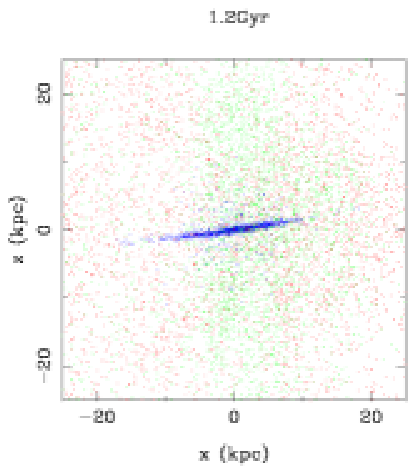}
	\hspace{.3cm}
	\includegraphics[angle=270,width=6cm]{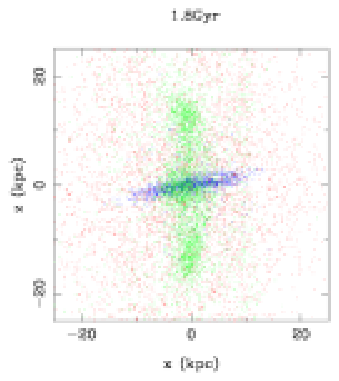}\\
	\vspace{.3cm}
	\includegraphics[angle=270,width=6cm]{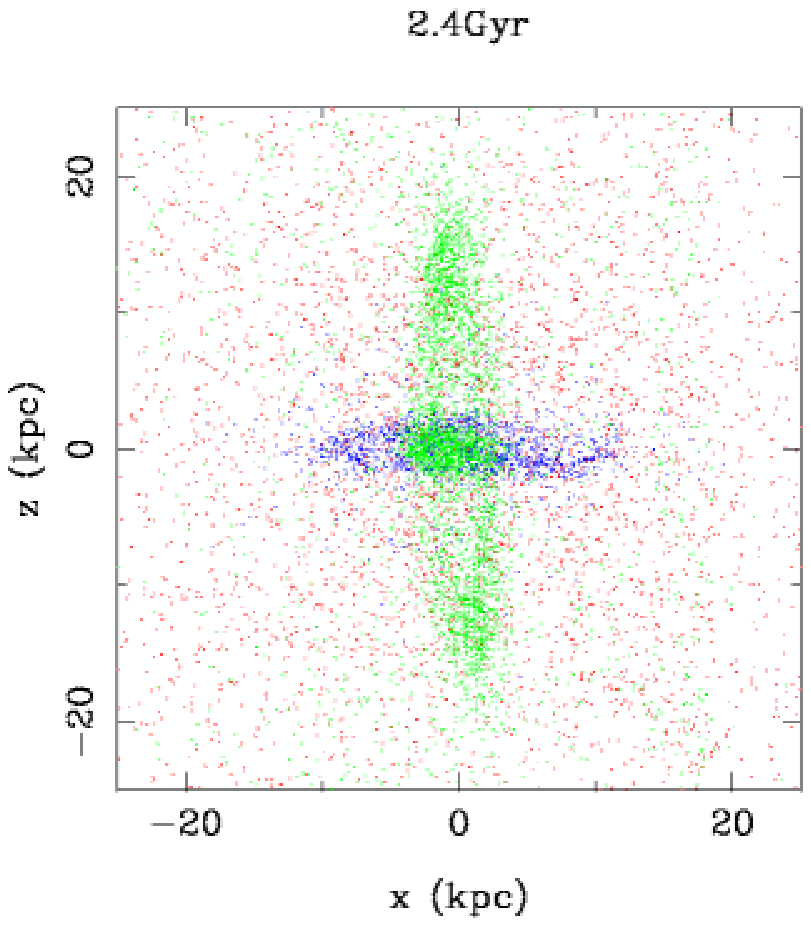}
	\hspace{.3cm}
	\includegraphics[angle=270,width=6cm]{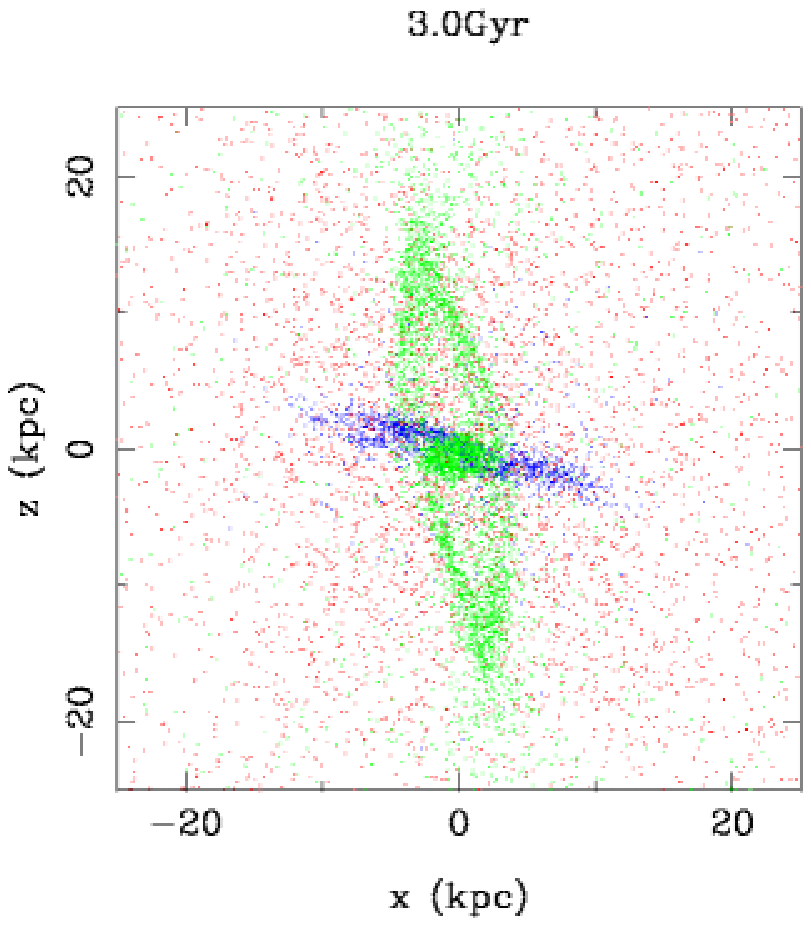}
	\caption{Particle distribution in run A9. Blue: intruder (stars) -- Red: victim (star) -- Green: victim (gas and stars formed from this gas after the ring).}
	\label{snapmerg}
	\end{figure*}

Our simulations confirm that the merging scenario provides an explanation for the formation of polar rings around pre-existing galaxies, and explains the various inner morphologies of polar rings. An example of galaxy merger that results in a polar ring is shown in Fig.~\ref{snapmerg} (run A9).

The main property of a polar ring, its radius, depends largely on the gas radial extent, and is only slightly influenced by other parameters. The potential of the host galaxy is thus not fundamental for the ring radius. This result is justified by an explanation of the ring formation in such galaxy mergers, given in Appendix~\ref{model}. The formation of very large rings ($R>10$ kpc) requires that the victim galaxy contains gas outside its stellar disk. Bekki (1998) explained the formation of large rings such as in NGC~4650A by low velocity collisions (about 30 km.s$^{-1}$), in which the victim disk is not dissolved, but such relative velocities are not common. It seems much more realistic to think that if polar rings are formed by the merging scenario large rings are related to victim galaxies that contain gas at large radii (Sancisi 1983), which is more likely than very low relative velocities.

When the mean radius of a polar ring $R$ is determined by Eq.~\ref{eqR}, we define its radial extent $\Delta R$ by:
 \begin{equation}
 \Delta R ^2=\frac{\int_0^\infty (r-R)^2 r \mu(r) dr}{\int_0^\infty r \mu(r) dr}
 \end{equation}
The values of $\Delta R/R$ that we obtain about 1 Gyr after the ring formation vary from 10\% to 40\%, and most of them are around 25\%.

	\subsubsection{Other components}~\label{oth_merg}
In every run, as in Fig.~\ref{snapmerg}, we notice the presence of:
\begin{itemize}
\item gas from the victim disk fallen on the center of the host galaxy
\item a stellar halo, formed by the stars of the victim that are dispersed by the collision
\end{itemize}

The fact that the stellar content of the victim remains dispersed was expected. Before the formation of the polar ring, gas is dispersed, and settles in a planar structure because of its dissipative nature. Stars are dispersed by the collision, too, but cannot settle in the polar plane. They then form a faint halo that surrounds the host galaxy and the polar ring. Runs in which the relative velocity exceeds 60 km.s$^{-1}$ show a nearly spherical stellar halo: Its axis ratio is larger than 0.75. When the relative velocity is smaller (which is much uncommon in the Universe), the victim galaxy stellar remnant is more flattened, yet remains much thicker than the gaseous ring: its axis ratio still exceeds 0.4, and differential precession makes it become nearly spherical in a few Gyrs. In 28 of our 34 runs ($>$ 80 \% of the runs), more than 50\% of the stellar mass of the victim is found within the radius of the polar ring. The presence of such a stellar halo is tested in observations of NGC~4650A (in Sect.~\ref{s6}). Let us keep in mind that the merging scenario results in a stellar structure that is distinct from the polar ring, even nearly spherical in most runs otherwise moderately flattened, and that more than half of the victim's stars are found within the polar ring mean radius $R$ in 80 \% of the simulations.

The polar ring does not contain all the gas of the victim's disk. A few percent remain dispersed at large radii, but the most important point is that some gas falls into the center of the host galaxy (see for instance Fig.~\ref{snapmerg}). This gas settles in an equatorial disk, but does not replenish the whole disk of the host galaxy, and never makes the host galaxy become gas-rich. Even if the polar ring is larger than the host galaxy, the gas fallen in the host galaxy remains located at small radii, generally not more than 2 or 3 kpc, and never more then 4 kpc in all our runs. This gas form a small central concentration in which it quickly forms stars, so that no gas remains after about 3 Gyrs. When the values of parameters are ideal ($\Theta$=$\Phi$=90 degrees, $R_{coll}$=0), the fraction of gas that falls in the center of the host galaxy is quite small, around 10-20 percent in most runs, at most 25 percent (run A8). This fraction becomes larger when we vary $\Theta$, $\Phi$, and $R_{coll}$ (see Sect.~\ref{constr_merg}). When the polar ring is formed, its angular extension is small enough to prevent differential precession to dissolve the ring and to make gas fall in the equatorial plane, but before the ring is formed, gas is dispersed with various inclinations to the polar axis, which enhances differential precession, and makes part of the gas fall in the host galaxy.

	\subsubsection{Ring inclination}\label{inclin_merging}
\begin{figure}
	\centering
	\includegraphics[angle=270,width=8cm]{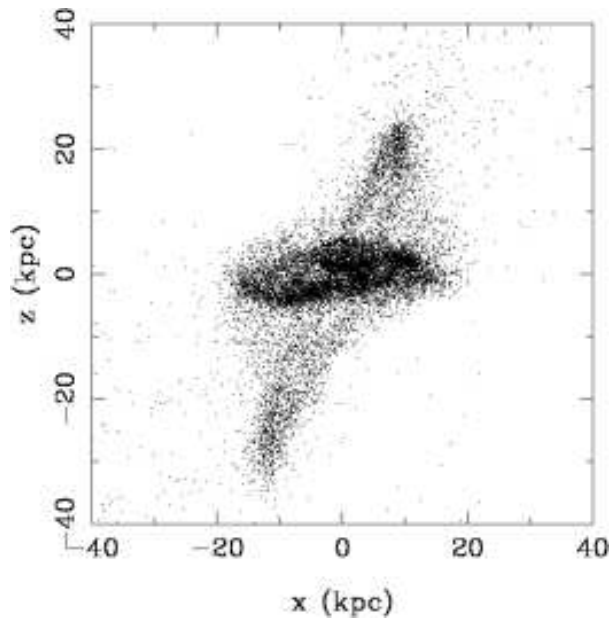}
\caption{Run B2: the polar ring mean inclination from the polar axis is 24 degrees, while the initial value of $\Theta$ was 65 degrees (i.e. 35 degrees from polar). Most polar rings formed in galaxy mergers are nearly polar, and no ring inclined by more than 25 degrees has been obtained. Stars from the victim disk that from a stellar halo are not represented.}
\label{incl}
\end{figure}

The inclination of the polar ring to the polar axis is influenced by the value of $\Theta$: small values of $\Theta$ result in rings that are inclined with respect to the polar axis. Yet, even if $\Theta$ is only 55 degrees, we obtain no ring inclined by more than 24 degrees from the polar axis (run B2, see Fig.~\ref{incl})

The final inclination of the polar ring is different from $\Theta$. We have mentioned that a significant fraction of gas falls inside the center of the equatorial disk. It takes always a polar component of the gas angular momentum (with respect to the host), so that the orientation of the mean angular momentum of the polar gas is modified. In other words, some gas from the ring falls inside the host disk, then the ring changes its orientation to conserve the total angular momentum. When $\Theta$ is 55 degrees, the resulting ring is closer to the polar axis, which explains why all the obtained rings are close to the polar axis

	\subsubsection{Stability of polar rings}\label{stabil_merging}
\begin{figure}
	\centering
	\includegraphics[width=8.5cm]{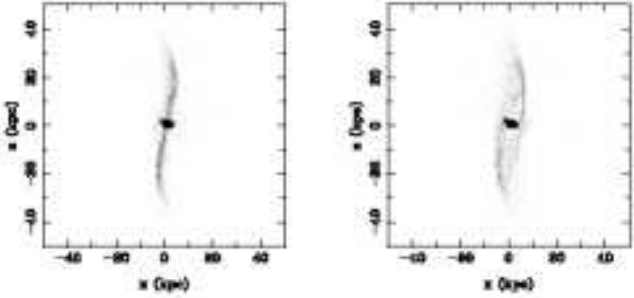}
	\caption{Gas and stars formed after the polar ring in run A11. The central condensation is related to gas that falls at the center of the host disk during ring formation and quickly forms stars. The polar ring is seen edge-on (left) and 10 degrees from edge-on (right).}
	\label{stabil}
	\end{figure}

Self-gravity is susceptible to stabilize polar rings (Sparke 1986, Arnaboldi \& Sparke 1994). Yet, the stability of polar structures in numerical simulations of formation should be addressed (Iodice et al. 2002a). Thus, we have let all the polar rings we have obtained evolve for at least 2 Gyrs. All were then stable. For some runs (A9, A11 and B2), we even let the system evolve for 8 Gyrs after the formation of the polar ring. Polar rings still appear to be stable. For instance, for run A11, we show the 8 Gyr old polar structure in Fig.~\ref{stabil}.

As displayed in Fig.~\ref{stabil}, most polar rings show a warp, that generally appears after 1 to 3 Gyrs of evolution. Theorical studies of the stability of self-graviting polar rings predict the presence of such warps (e.g. Arnaboldi \& Sparke 1994): self-gravity counteracts differential precession through the warp. 

No unstable polar ring has been observed in the merging scenario. As explained in Sect.~\ref{inclin_merging}, rings are always nearly polar, and their inclination from the host plane always exceeds 75 degrees. It seems that the polar ring forms only if its inclination is large enough to allow self-gravity to counteract differential precession. Before the formation of the polar ring, gas is dispersed around the host galaxy. Dissipation and self-gravity tend to make it gather in a polar plane, while differential precession is already acting. Two situations may then occur: either the differential precession is too large to be counteracted by gravity and no ring forms, or the differential precession is small enough to ensure that the ring will both form and be stable. This simply explains why all the polar rings obtained in the merging scenario where stable. We will see a different situation for the accretion scenario (Sect.~\ref{s4}), in which rings may form and be unstable.

In Fig.~\ref{stabil}, we also notice that:
\begin{itemize}
\item the polar ring is not totally depleted, and the central hole becomes less obvious after several Gyrs. Star formation is responsible for such a dispersion of the polar matter: we have not obtained this result in the same run without star formation. 
\item in some lines-of-sight, the ring shows two spiral arms, while no real arms are seen in the face-on ring. These spiral arms are actually related to the warp, and are observed when the polar ring is seen nearly edge-on, but not perfectly edge-on. For instance, the spiral structures observed in NGC~4650A (Arnaboldi et al. 1997) may not be density waves but be caused by a warp. 
\end{itemize}

\subsection{Constraints on parameters} \label{constr_merg}
In this part, we investigate the influence of deviations from the ideal scenario. We wish to show in which range each parameter can be varied without preventing the formation of a polar ring. Thus, we examine the results of runs B1 to B21. As explained in Sect.~\ref{inclin_merging} and ~\ref{stabil_merging}, polar rings form in a merging event only if the effects of differential precession are small enough, which requires that gas initially orbits on nearly-polar paths. When we deviate from this condition, the larger differential precession makes more gas fall to the center of the host galaxy. If the deviation is too large, all the gas falls inside the host galaxy and no polar ring is obtained. 

In runs A1 to A13, the fraction of the victim's gas that falls to the center of the host galaxy does not exceed 25 percent. We consider that the effects of varying a parameter become significant when this fraction is between one and two third, and that a ring was bound not to form if this fraction exceeds two thirds. This is justified by the fact that a large fraction of gas fallen on the host center attests to a nearly unstable ring (because of large differential precession effects), and a nearly unstable ring is likely not to form at all (see Sect.~\ref{stabil_merging}: no unstable rings are obtained in this scenario). We then describe the results of runs B1 to B19 in Table~\ref{res2} in terms of ``ring'' (the variation of parameters from their ideal value is not significant), ``light ring'' (the variation of parameters becomes significant), ``very light ring'' (much matter is lost by the polar ring, which is thus nearly not formed), and ``no ring'' (the variation in parameters is too large). 

    \begin{table}
 	   \centering
           \begin{tabular}{cl}
           \hline
	   \hline
           Run & result \\
           \hline
           B1  & ring            \\
           B2  & ring            \\
           B3  & no ring         \\
           B4  & ring            \\
           B5  & no ring         \\
           B6  & ring            \\
           B7  & very light ring \\
           B8  & ring            \\
           B9  & ring            \\
           B10 & light ring      \\
           B11 & no ring         \\
           B12 & ring            \\
           B13 & light ring      \\
           B14 & very light ring \\
           B15 & very light ring \\
           B16 & no ring         \\           
           B17 & ring            \\
           B18 & light ring      \\
           B19 & no ring         \\
           B20 & very light ring \\
           B21 & no ring         \\
           \hline
           \end{tabular}
           \caption{Results of runs B1 to B21. ``light ring'': a large part (more than one third) of the victim's gas falls inside the center of the host galaxy and is lost by the polar ring during its formation. ``very light ring'': less than one third of the victim's gas is found in the polar ring after its formation, which indicates a large rate of differential precession: such rings are nearly unstable, thus likely not to form, since no unstable ring forms in the merging scenario (Sect.~\ref{stabil_merging})}\label{res2}
           \end{table}

The results of runs B1 to B21 then allow us to determine the constraints on each parameter, i.e. the range in which each parameter may be varied without preventing a polar ring from forming. These constraints are:

\begin{itemize}
\item $\Theta$ $>$ 50 degrees
\item $\Phi$ $>$ 45 degrees
\item The victim galaxy should contain 3 times more gas than the intruder, or have twice as large a gaseous disk, in order to prevent all the gas from merging in a single, equatorial disk
\item $R_{coll}$ $<$ 0.5, in units of the victim disk radius
\item The relative velocity should be small enough to allow the two systems to merge, and should typically not exceed 150-200 km.s$^{-1}$. The larger the velocity, the larger the constraints on other parameters (see runs B1-2-3 vs. runs B9-10-11).
\end{itemize}

Each parameter can be significantly varied, thus the merging scenario is actually robust, and may explain the formation of a significant number of polar rings. We will compare constraints on both scenarios and determine which is the most likely to occur in Sect.~\ref{s5}.

\subsection{Another model for the merging scenario}

A semi-analytical model for the merging scenario is presented in Appendix~\ref{model} to explain why a ring forms in this scenario. The main results of this model are:
\begin{itemize}
\item Polar gas oscillates at large and small radii before gathering in the ring. It is not directly placed in the ring, for the galaxy encounter stimulates large radial motions. In particular, even if the ring is larger than the host disk, polar gas will first encounter the host galaxy material
\item The ISM dissipative nature is responsible for the formation of a polar ring rather than a polar disk. 
\end{itemize}
Detailed explanations of these results are given in Appendix~\ref{model}

\section{Accretion scenario}\label{s4}
\subsection{Relevant parameters and numerical simulations}

	\begin{table*}
	\centering
	\begin{minipage}{18cm}
	\centering
	\begin{tabular}{ccccccccccccc}
	\hline
	\hline
Run & $m^G_\mathrm{host}$ & $r^G_\mathrm{host}$ & $m^G_\mathrm{donor}$ & $r^G_\mathrm{donor}$  &  $V$ &  $\Theta$ &  $\Phi$ &  $R_m$ \\
\footnote{(e): elliptical host galaxy}&
\footnote{host gas mass fraction}&
\footnote{host gas radial extent, in units of stellar radius}&
\footnote{donor gas mass fraction}&
\footnote{donor gas radial extent, in units of stellar radius}&
\footnote{relative velocity before the encounter}&
\footnote{angle between the donor disk plane and the host equatorial plane}&
\footnote{angle between the donor orbital plane and the host equatorial plane}&
\footnote{minimal radius between the host center and the donor center, in units of the host optical radius}\\
\hline
C1  & 0 & - & 0.3 & 3.0 & 200 & 90 & 90 & 5  \\ 
C2  & 0 & - & 0.3 & 3.0 & 300 & 20 & 90 & 5  \\ 
C3  & 0 & - & 0.3 & 3.0 & 350 & 90 & 60 & 5  \\ 
C4  & 0 & - & 0.2 & 3.0 & 200 & 90 & 90 & 5  \\ 
C6  &0.2& 1.& 0.3 & 1.5 & 150 & 80 & 80 & 5  \\ 
C7  & 0 & - & 0.2 & 2.5 & 200 & 90 & 50 & 6  \\ 
C8  & 0 & - & 0.4 & 2.5 & 200 & 90 & 50 & 6  \\ 
C10 & 0 & - & 0.3 & 2.5 & 250 & 70 & 70 & 6  \\ 

C11 & 0 & - & 0.35& 5.0 & 230 & 80 & 80 & 6  \\ 
C12 & 0 & - & 0.5 & 6.0 & 250 & 65 & 65 & 8  \\ 

C16(e)& 0 & - & 0.3 & 5.0 & 175 & 90 & 90 & 6  \\ 
C17(e)& 0 & - & 0.3 & 5.0 & 150 & 90 & 75 & 4  \\ 
C18(e)& 0 & - & 0.2 & 4.0 & 250 & 90 & 75 & 8  \\ 
C19(e)& 0 & - & 0.3 & 5.0 & 175 & 40 & 25 & 4.5  \\ 
C20(e)& 0 & - & 0.2 & 4.0 & 250 & 25 & 40 & 5  \\ 
C21(e)& 0 & - & 0.2 & 4.0 & 250 & 10 & 10 & 5  \\ %

C22 & 0 & - & 0.2 & 4.0 & 250 & 45 & 20 & 5  \\ 

C23 & 0 & - & 0.3 & 4.0 & 200 & 70 & 55 & 7  \\ 
C24 &0.25&1.& 0.3 & 4.0 & 200 & 65 & 55 & 6  \\ 
C25 & 0 & - & 0.3 & 4.0 & 200 & 80 & 65 & 5  \\ 
C26 &0.3&1.4& 0.3 & 4.0 & 200 & 65 & 65 & 5  \\ 

C27 & 0 & - & 0.5 & 4.0 & 100 & 90 & 90 & 0  \\ 

	\hline
	\end{tabular}
	\caption{Run parameters. Units: $10^9$ M$_{\sun}$, kpc, km.s$^{-1}$}\label{par2}
	\end{minipage}
	\end{table*}

As was the case for the first scenario, some numerical tests have proved that inner parameters of both galaxies (disk, bulge and halo masses, disk radius) are not the most relevant. The most important parameters, given in Table~\ref{par2}, are the gas content (mass and radial extent) of each galaxy, and orbital parameters, which are:
\begin{itemize}
\item the relative velocity $V$ before the encounter. To deduce the velocity at the beginning of the simulation, we neglect effects of dynamical friction, and treat galaxies as two rigid bodies for which the conservation of kinetic energy is computed.
\item the angle $\Theta$ between the donor disk plane and the host equatorial plane, given before the accretion event (tidal effects may make it evolve).
\item the angle $\Phi$ between the donor orbital plane and the host equatorial plane.
\item the minimal radius $R_m$ between the host center and the donor center, in units of the host optical radius. In run C27, this value is zero, for both galaxies merge after the ring accretion (notice that the ring is not issued from the merger as was the case in the previous scenario, but really formed by tidal tails before the merging of the donor and the host).
\end{itemize}

For runs C16 to C21, the host galaxy is an E4 elliptical. In fact, even if the host stellar distribution is changed from a spiral disk, the large-scale potential is not fundamentally different, and we notice no major modification in the evolution of the donor and the accreted gas.

\subsection{General properties}
	\subsubsection{Morphology of the polar ring}~\label{morph_accr}
In Table~\ref{res_accr}, we describe the result of each run.

	\begin{table}
	\centering
	\begin{tabular}{cc}
	\hline
	\hline
Run & result \\
\hline
C1  & polar ring \\
C2  & polar ring (25 degrees from polar)\\
C3  & polar ring \\
C4  & no ring (no accretion)\\
C6  & polar ring \\
C7  & polar ring \\
C8  & polar ring (20 degrees from polar)\\
C10 & polar ring  \\ 
C11 & polar ring (see Fig.~\ref{accret4650})  \\ 
C12 & polar ring  \\ 
C16(e)& polar ring\\
C17(e)& polar ring\\
C18(e)& polar ring\\
C19(e)& polar ring (25 degrees from polar) \\
C20(e)& inclined ring (30 degrees from polar) \\ 
C21(e)& equatorial dust lane \\
C22 & polar ring (25 degrees from polar)   \\ 
C23 & polar ring \\
C24 & polar ring \\
C25 & inclined ring (40 degrees from polar) \\
C26 & inclined ring (40-45 degrees from polar) \\ 
C27 & polar ring, then polar disk \\ 
	\hline
	\end{tabular}
	\caption{Run results}\label{res_accr}
	\end{table}

\begin{figure*}
	\centering
	\includegraphics[angle=270,width=6cm]{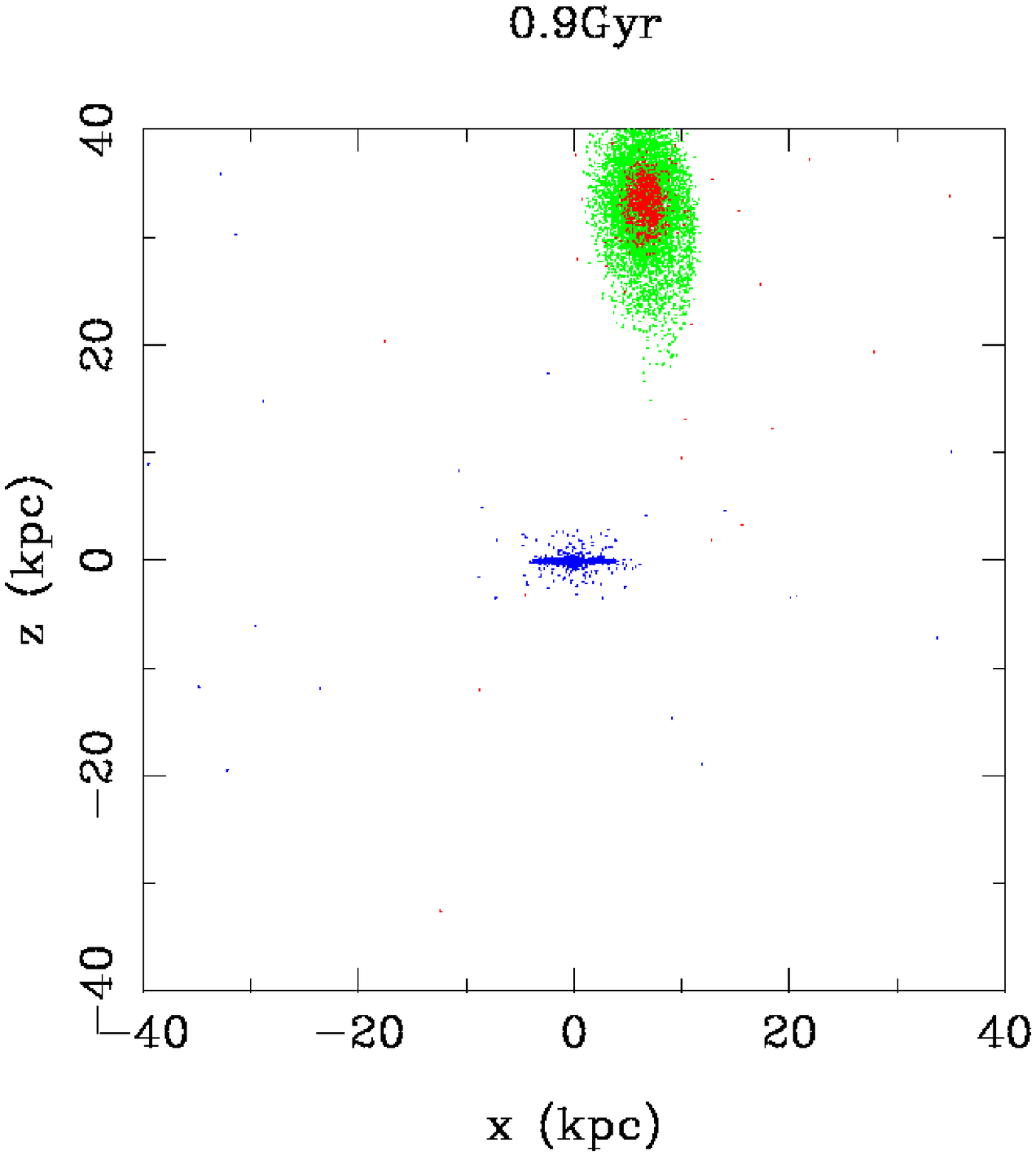}
	\hspace{.3cm}
	\includegraphics[angle=270,width=6cm]{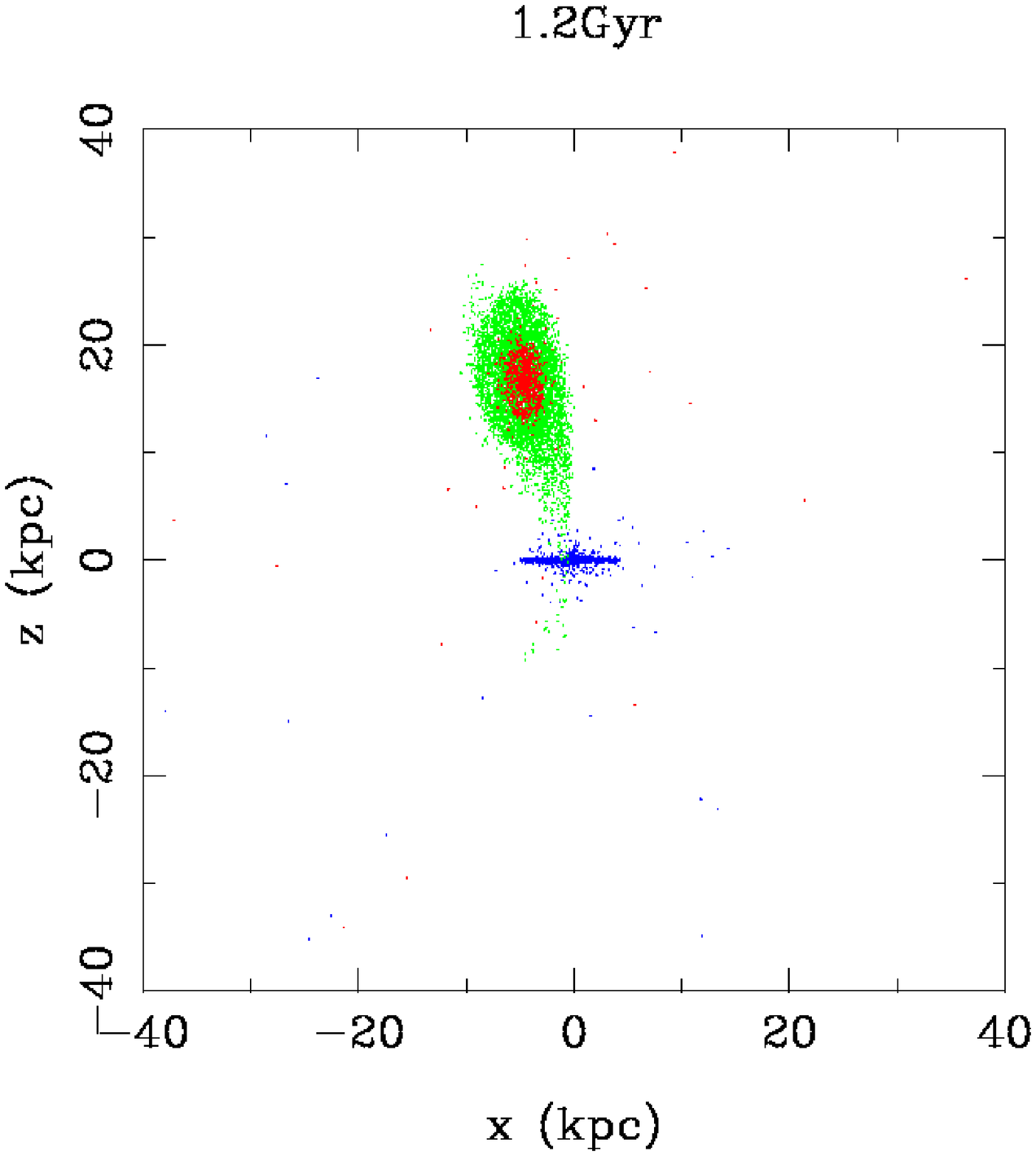}\\
	\vspace{.3cm}
	\includegraphics[angle=270,width=6cm]{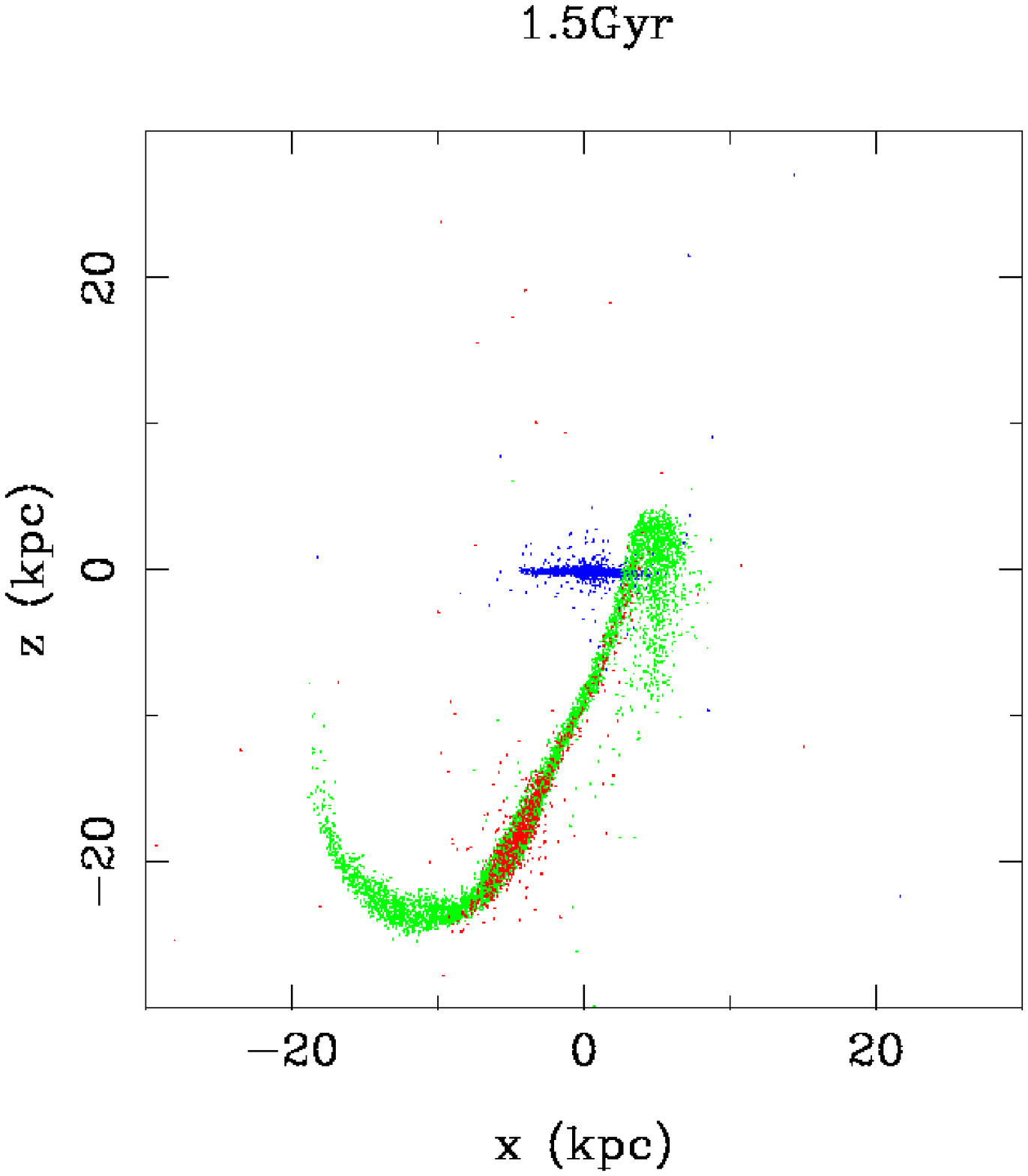}
	\hspace{.3cm}
	\includegraphics[angle=270,width=6cm]{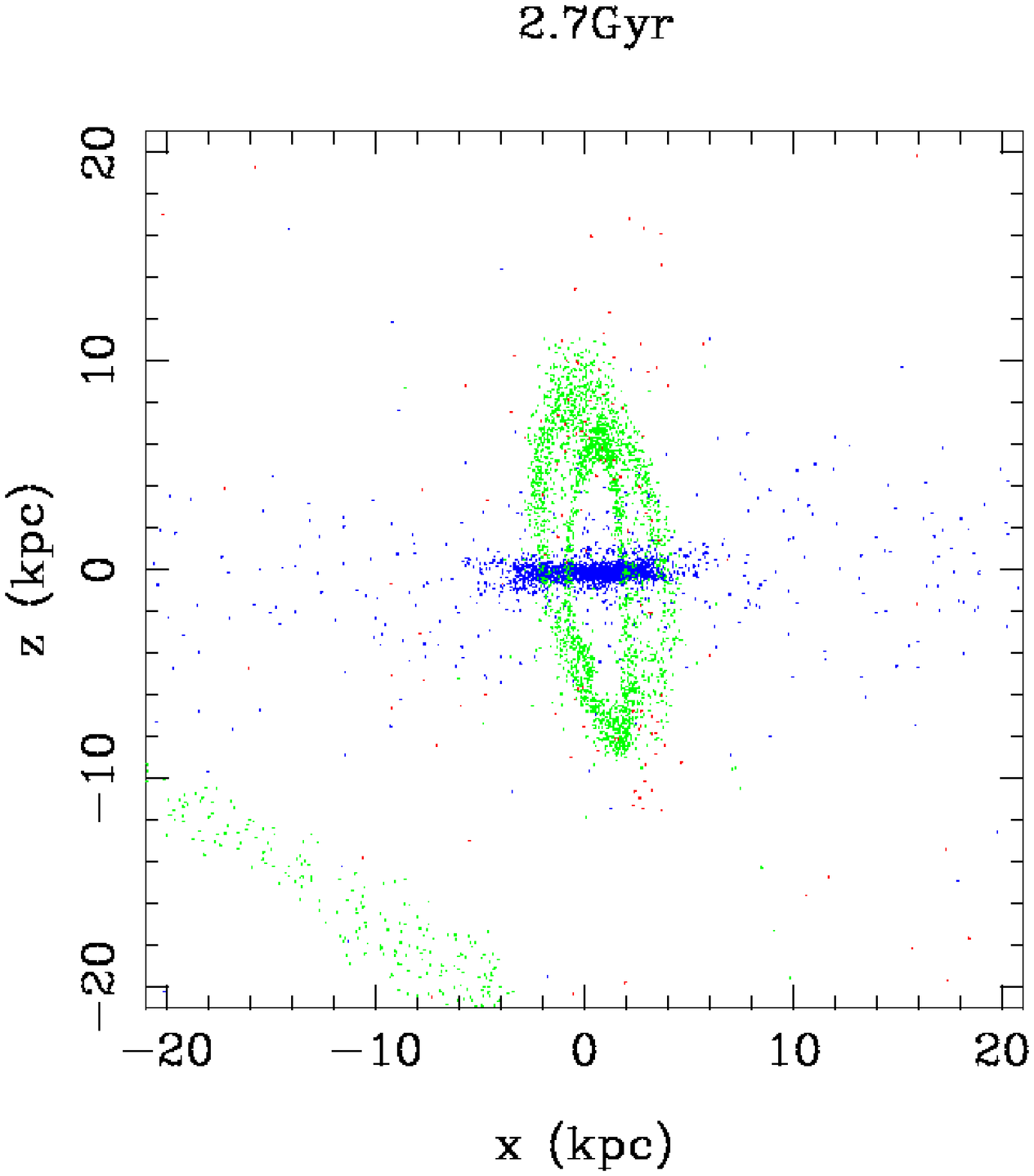}\\
	\vspace{.3cm}
	\includegraphics[angle=270,width=6cm]{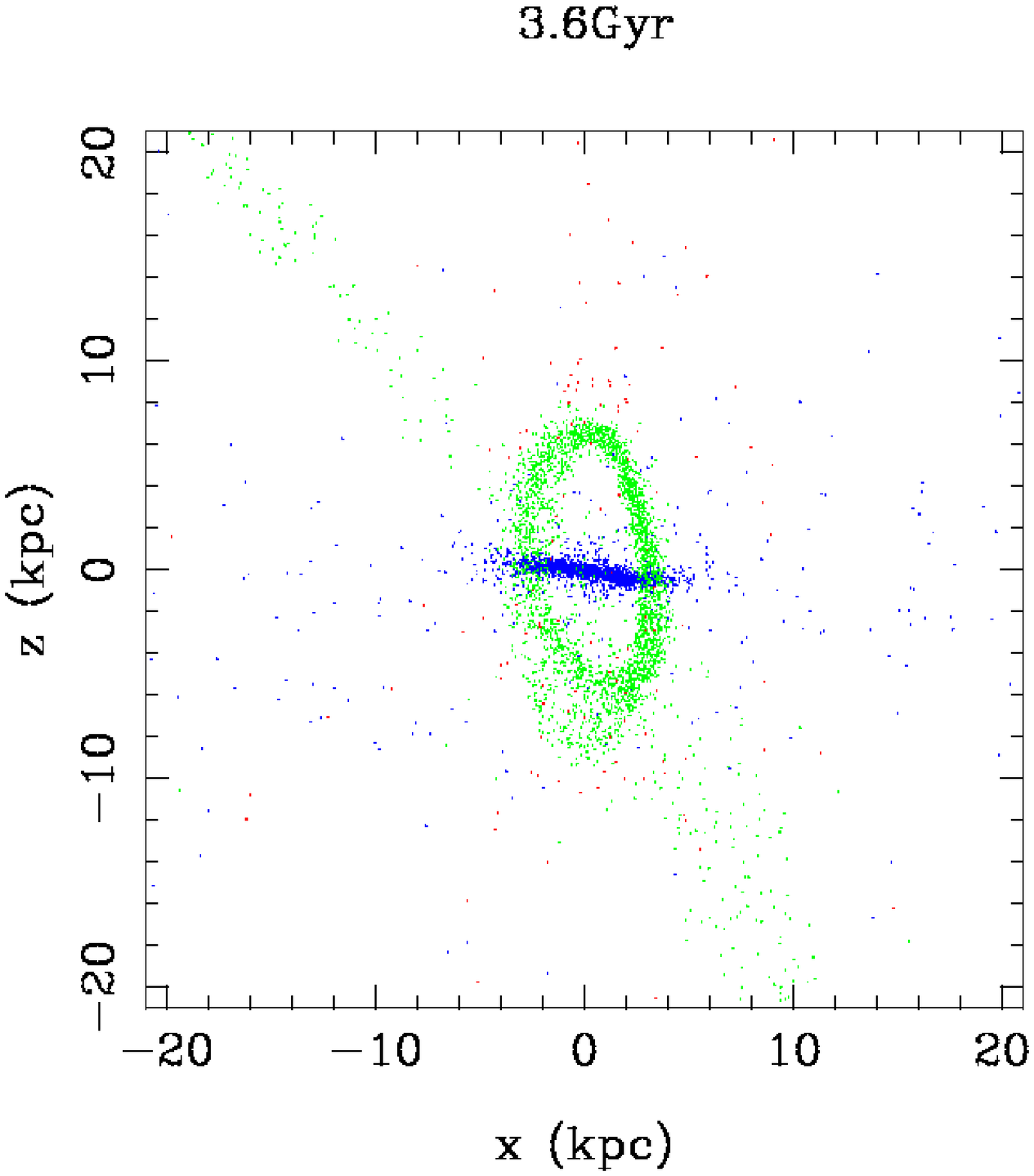}
	\hspace{.3cm}
	\includegraphics[angle=270,width=6cm]{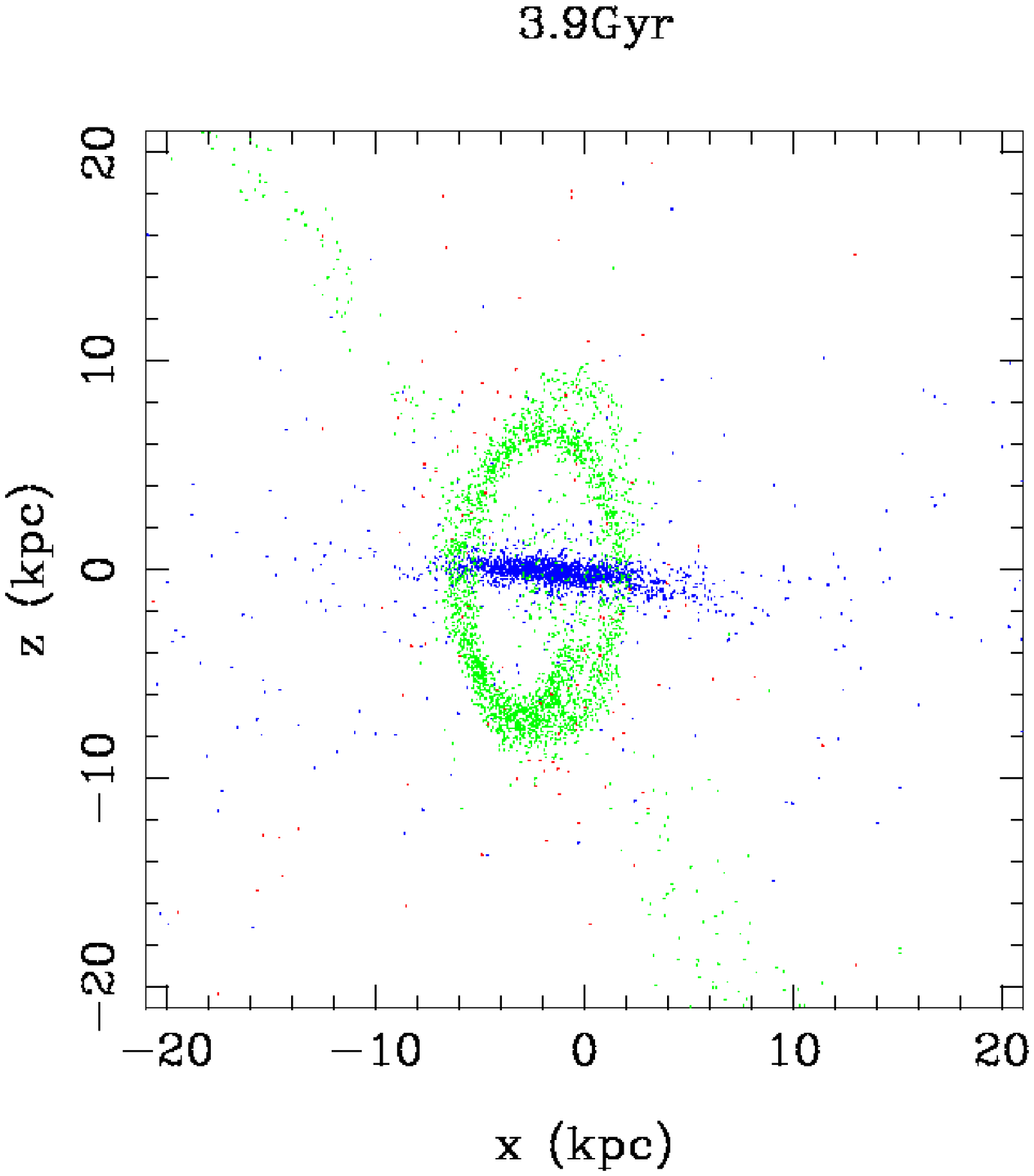}
	\caption{Particle distribution in run C1. Blue: host (stars) -- Red: donor (star) -- Green: victim (gas and stars formed after the ring in this gas). }
	\label{snap_accret}
	\end{figure*}

An example of formation of a polar ring by tidal accretion is shown of Fig.~\ref{snap_accret} (run C1). As in the merging scenario, tidal accretion accounts for a large variety of polar structures. The main property of a polar ring, its mean radius $R$, may be smaller as well as larger than the host galaxy optical radius, as already shown by Reshetnikov \& Sotnikova (1997). They also found that the host dark halo influences the value of $R$, which some of our simulations confirm. As well as the merging scenario, accretion produces warped rings and rings with inner spiral structures ; transient helicoidal features resulting in polar rings are also observed. 

\begin{figure}
	\centering
	\includegraphics[angle=270,width=8cm]{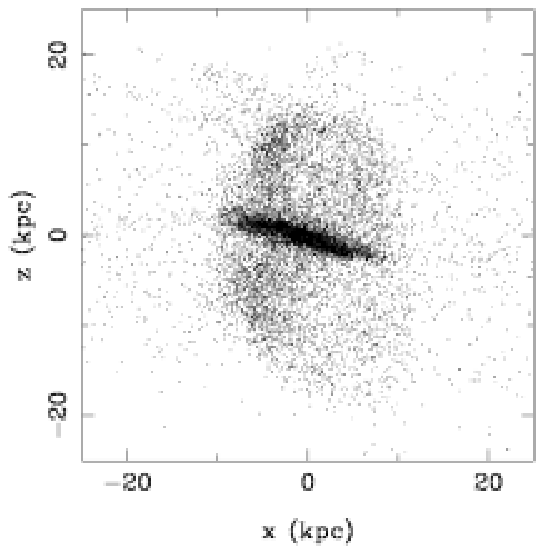}
	\caption{Run C27: the merging of a part of the donor galaxy with the host galaxy, after the formation of the polar ring, makes the radial extent of the polar ring raise to 55\%, so that this structure could be regarded as a polar disk.}
\label{polardisk}
\end{figure}

Just after their formation, rings have a small radial extent: $\Delta R/R$ varies from 10\% to 35\%. $\Delta R$ seems at this stage smaller than in the merging scenario. Yet, in run C27, the merging of the donor galaxy with the host galaxy, after the accretion of the polar ring, disturbs the ring, and makes $\Delta R/R$ raise from 25\% to 55\% (see Fig.~\ref{polardisk} and discussion in Sect.~\ref{diskvsring}). More generally, rings may be disturbed by interactions with other galaxies, which will have a similar effect: even if rings have a small radial extent just after their formation in this scenario, they may get a larger radial extent later.

\begin{figure}
	\centering
	\includegraphics[angle=270,width=8cm]{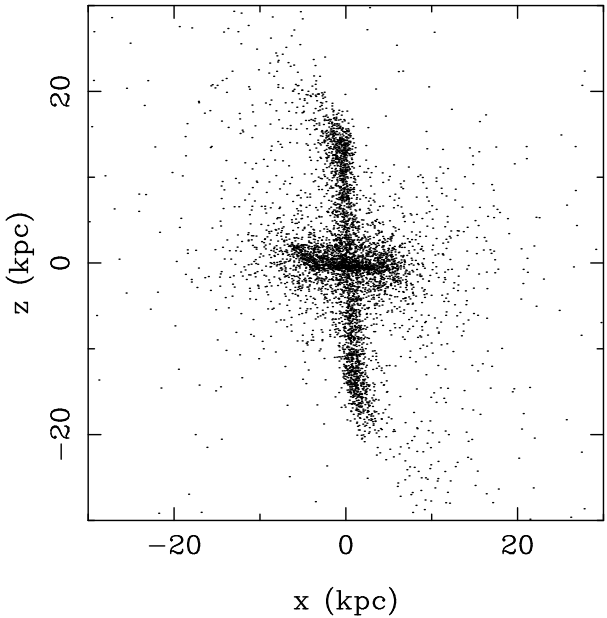}
	\caption{Run C11: a very massive polar ring is formed by tidal accretion, the ring mass is about 80\% of the host mass. The ring is slightly warped in its outer regions.}
\label{accret4650}
\end{figure}

The last noticeable point concerns the mass of polar rings. One could a priori think that:
\begin{itemize}
\item for gas to be accreted from the donor by the host galaxy, the donor must be less massive than the host
\item only a small fraction of the donor is accreted
\end{itemize}
This led Iodice (2001) and Iodice et al. (2002a) to say that rings formed by accretion would be significantly lighter than host galaxies, unlike some observed rings as in NGC~4650A. Actually, the fraction of accreted gas may reach 40\% of the donor gaseous mass (run C11), provided that the gas extent in the donor is large enough (gas at large radii is not strongly gravitationally bound) and the donor inner rotation is favorable to accretion: the accreted mass can be large in the case of direct resonance (run C3) while no gas is accreted in run C4 (retrograde resonance). Moreover, if the donor contains gas outside this stellar disk, accretion can be achieved from the donor to the host even if the donor is more massive than the host (for instance run C8 where the donor mass is 4 times larger than the host mass, and accretion still occurs from the donor to the host). Finally, with large donor masses and large accreted fractions, the ring masses are sometimes of the same order as the host mass (see Fig.~\ref{accret4650}). An observational example of a galaxy that is accreting a large gas mass from a more massive galaxy is NGC~646 (Danziger \& Schuster 1974, Horellou \& Booth 1997).

	\subsubsection{Other components}
In the merging scenario, we have shown that the host galaxy had to contain little gas. Here, the host galaxy may be gas rich: in runs C24 and C26, the host gas mass fraction is larger than 25\% before the accretion event, and still larger than 20\% 1 Gyr after the ring formation (see also Sect.~\ref{obs660}). During the formation of the polar ring, accreted matter is placed directly around the ring radius $R$, thus gas present in the host galaxy at radii larger or smaller than $R$ will not disturb the ring formation, for both gaseous component will not meet. In the merging scenario, polar gas oscillates at different radii because of the large change in the central mass (see Appendix~\ref{model}) and always encounters the intruder gas, so that equatorial gas at some radii is much more able to prevent the polar ring from forming, even if present only at radii smaller than the final ring radius $R$.

Another important difference to the first scenario is that stars are generally not accreted from the donor, for they form a warm, non-dissipative, medium. The fraction of accreted stars in our simulations is always smaller than 10\% and nearly always smaller than 5\%. Moreover, the few accreted stars join the polar ring via the same tidal tail as accreted gas. Thus, no stellar halo is expected around the polar ring, contrary to the merging scenario (see Sect.~\ref{oth_merg}).
	
	\subsubsection{Stability and inclination of polar rings}\label{stab_accret}
The stability concern does not rule out the accretion scenario : 
polar structures are as stable as in the merging scenario 
(some rings were simulated over 10 Gyrs), 
except some highly inclined rings that are dissolved in a few Gyrs (see 
Sect.~\ref{obs660}). The relevant point is not that rings formed by 
accretion are less stable due to their formation history, but that they 
are sometimes more inclined. In the merging scenario, the final ring was 
closer to the polar axis than the initial victim galaxy, because of the 
infall of polar gas inside the center of the equatorial system during the 
ring formation. In the accretion scenario, we do not observe such a 
systematic large gas infall, yet the ring orientation is still 
different 
from the initial value of $\Theta$. The likely explanation is that only 
a part of gas is found in the polar ring: some gas is sometimes found in 
the host galaxy (but this effect is not as large and as general as 
in the merging scenario), some gas remains in the donor, and some gas is 
dispersed at 
large radii from both galaxies. The polar ring gas has only a part of the 
initial angular momentum of the donor gas, thus is not compelled to be 
parallel to the donor initial disk. For instance, this explains why two 
nearly parallel disks, or disks on a nearly equatorial orbit, may result 
in a nearly polar ring: such unexpected situations are observed in runs 
C19,20,22. In these runs, initial parameters are far from ideal values for 
the formation of a polar structures, but the final ring is not far from 
polar. In run C22, the ring is about 25 degrees from polar, while 
initially $\Theta=45$ degrees. This allows this ring to be stable (it has 
been followed in the simulation during 4 Gyrs after its formation). 

With the same inclination (i.e. less than 25 degrees from the polar axis), rings are as stable as in the merging scenario, and the stability concern does not seem to be closely linked to the formation mechanism. In the accretion scenario, we find in addition more inclined rings. This will be discussed in Sect.~\ref{obs660}.

\subsection{Constraints on parameters}\label{constr_accr}

We will see in Sect.~\ref{prob_accr} that parameters cannot be regarded as independent in this scenario. Thus it is not as easy to determine a limit on each parameter as it was in the merging scenario. However, doing this is not required in the forthcoming comparison of scenarios. We will simply need to know in Sect.~\ref{prob_accr} that constraints on parameters in this scenario are significantly smaller than in the merging scenario: 
\begin{itemize}
\item we observe the formation of polar rings with values of $\Theta$ and $\Phi$ smaller than in the merging scenario, as explained in Sect.~\ref{stab_accret}
\item relative velocities and distance between galaxies can be rather large without preventing a massive ring from being accreted (for instance run C11)
\item host galaxy may contain a significant fraction of gas (runs C24, C26) ; this will be largely discussed in Sect.~\ref{obs660}. Before the ring formation, the host galaxy may even be gas rich and contain gas outside its stellar disk (run C26).
\end{itemize}
The three categories of parameters (gas content of the host galaxy, orientation of galaxies, relative velocities and radius) are thus submitted to smaller constraints than in the merging scenario. This will now be the issue of the first comparison of both scenarios.

\section{Probability of each scenario and abundance of polar rings}\label{s5}
\subsection{Probability of the merging scenario}
In Sect.~\ref{constr_merg}, we established in what conditions the merging of two spiral galaxies results in a polar ring. We now derive the associated probability: we assume that two spiral galaxies collide and merge, and determine the probability that this event forms a polar ring:

\begin{itemize}
\item the probability that $\Theta$ $>$ 50 degrees is 0.45
\item the probability that $\Phi$ $>$ 45 degrees is 0.29
\item the probability that $R_{coll}$ $<$ 0.5 is 0.25
\item the constraint on the gas content is applied to HI observations of spiral galaxies (Broeils \& Rhee, 1997): we select randomly two galaxies, and check whether one of them may be a victim galaxy, i.e. contain three times more gas then the other one, or have a twice as large gaseous disk. The probability that a randomly chosen pair of galaxies is a possible intruder-victim pair, without regarding which one is the victim, is 0.3.
\end{itemize}

The probability that two merging spirals form a PRG is then $0.25\times0.29\times0.3\times0.45=1\%$.

The constraint on the relative velocity is more difficult to convert into probability. Yet, it seems reasonable to admit that a given galaxy has undergone at most 5 merging events with spiral galaxies (the so-called ``major mergers''). Merging events with dwarf galaxies are ignored, for they would not account for the large masses of the observed polar rings. The adopted value of 5 major mergers is a large upper limit, and at each merging event, the probability that a polar ring forms is 1\%. Finally, the probability that a possible host galaxy (lenticular or early-type spiral) has acquired a polar ring during a merging event is {\it at most} 5\%.

The simulation parameter choices that we have made (see Table \ref{par}) may seem arbitrary: for instance, the value of $\Theta$ is varied in more runs than the value of $\Phi$, and some situations have not been studied (for instance $\Theta=45$ and $\Phi=45$ degrees in the same run). The statistics based on simulations may then seem biased, but is actually not. The presented runs have been used only to delimit the range of parameters for which a polar ring is formed in agreement with observations. From the obtained range of values for each parameter, we can determine the probability represented by these combined ranges, and derive the statistics.

\subsection{Probability of the accretion scenario}\label{prob_accr}
In the merging scenario, we have established independent constraints on each parameter, in order to determine the probability that a polar ring is formed during a merging event. This cannot be done so easily for the accretion scenario, for:
\begin{itemize}
\item the frequency of galaxy interactions is not known, while we had been able to fix an upper limit to the frequency of major mergers.
\item the parameters for the accretion scenario are not independent. For instance, the sense of rotation of the donor is crucial, so that two perpendicular disks on a polar orbit may form no ring in some situations, while two parallel disks on a nearly equatorial orbit may form a polar ring in some other situations. 
\end{itemize}
As parameters cannot be regarded as independent, deriving a dependable probability for this scenario would require us to vary all the relevant parameters in the same time, while we had varied one parameter after the other for the merging scenario. A large number of simulations would then be required, but deriving a precise probability for the accretion scenario is not required for our study, for previous results allow us to qualitatively compare the probability of both scenarios: we have simply shown that polar rings can form by accretion in a large range of parameters, with values of $\Theta$ and $\Phi$ that are far from their ideal values. The constraints are smaller than for the merging scenario: two perpendicular disks have been observed to form a polar ring by tidal accretion, while they never form a polar ring according to the merging scenario. 

Not only are the constraints on parameters smaller, but also tidal interaction events (that may lead to mergers or not) are much more frequent than galaxy collisions leading to mergers, for the radius between galaxies may be much larger and a large relative velocity does not prevent tidal accretion of much matter (we have obtained 40\% of accreted gas in run C11, with $V=230$ km.s$^{-1}$ and $R_m=6$). Thus, the probability that a possible host galaxy has acquired a polar ring by tidal accretion is significantly larger than the probability that it has acquired a polar ring during a merger event. 

\subsection{Abundance of polar rings and comparison of scenarios}
From their atlas of PRGs, Whitmore et al. (1990) estimate that 4.5\% of the possible host galaxies have acquired a polar ring, and say that this estimate is uncertain by about a factor 3. 

According to the merging scenario, polar rings are stable features, so that no polar rings would have disappeared. 1.5\% to 13.5\% of the possible host galaxies should then have a polar ring. This is obviously a lower limit, for many rings may remain unknown. On the other hand, we have shown that at most 5\% of the possible host galaxies have acquired a polar ring during a merging event. Thus, as 5 is in between 1.5 and 13.5, we can claim that:
\begin{itemize}
\item the probability of the merging scenario is large enough to account for all the observed polar rings
\item the probability of the merging scenario would become too small, if many polar rings have not been detected yet. 
\end{itemize}

We have not determined a probability for the accretion scenario, but explained that it was much more probable than the merging scenario. Thus:
\begin{itemize}
\item the probability of the accretion scenario is large enough to account for all the observed polar rings
\item the probability of the accretion scenario would still be large enough, if many polar rings have not been detected yet. The large probability of the accretion scenario even implies that many polar rings have not been detected yet, for they are for instance not massive enough.
\end{itemize}

Probability concerns do not allow us to rule out one of the scenarios. 
Even if the probability of the merging scenario is much smaller than the probability of the accretion scenario, only discoveries of many new PRGs, that would increase the observed fraction of polar rings, would weaken this scenario.

\section{Comparison with observations}\label{s6}
\subsection{General considerations}
In Sect.~1, we mentioned several observational properties of polar rings. Most of them are reproduced by both formation scenarios:
\begin{itemize}
\item most host galaxies are early-type, which is explained by the constraint on the gas content of host galaxies, related to the fact that two orthogonal gaseous systems cannot be found at the same radius.
\item polar rings may be very massive
\item polar rings have various radii
\item most polar rings are nearly polar
\item most polar rings(even all rings in the merging scenario) are very stable and may be observed after several Gyrs
\item polar rings show various inner morphologies
\end{itemize}
The remaining properties may discriminate between scenarios: they are the existence of gas-rich host galaxies, of inclined rings, and the possible presence of a stellar halo around the polar ring (Sect.~\ref{oth_merg}). We now compare these properties in simulations and observations.

\subsection{NGC~4650A}
	\subsubsection{Photometry}

The main photometric difference between the numerical results of both scenarios is the presence of a stellar halo in which the polar ring is embedded according to the merging scenario. This stellar halo is the remnant of the stellar component of the victim galaxy, which is dispersed during the merger: only gas settles in a polar plane, due to its dissipative nature. According to the accretion scenario, only gas is accreted, for it is colder than stars and is generally present at larger radii in the donor galaxy. Nearly no stars are accreted from the donor galaxy, even if the donor galaxy has a massive stellar disk. Obviously, stars are found in the polar ring because of star formation, but no stellar halo is obtained. Testing the presence of a stellar halo around a PRG may then give insight into its history. 

We have generally observed in simulations of merging events that most of the stellar mass of the victim settles inside the ring radius $R$. More precisely, we have noticed that in 80\% of our simulations, more than 50\% of the victim's stellar mass is observed inside the disk radius $R$. 

We use deep HST observations of NGC~4650A (Gallagher et al. 2002) and try to detect a stellar halo. Optical observations have already shown some stars rather far from the polar ring plane (see Fig.~12 in Arnaboldi et al. 1997), yet it seems premature to attribute them to a stellar halo, for HI gas (that belongs to the polar ring) is observed at the same locations (see Fig.~2 in Arnaboldi et al. 1997). Deep HST observations also find stars at large radii, but a detailed analysis is required to decide whether they belong to the faintest regions of the polar ring and the host galaxy or form a third component.

Iodice et al. (2002a) have shown that the structure of NGC~4650A is well described, even at large radii, by a two-component 
model. This suggests that no stellar halo is required to account for the observed luminosity. Yet, we still do not know if really no halo is present, or if the halo is too faint to be detected.

To estimate this, we build a luminosity model that accounts for the host galaxy and the polar ring. The model for the host galaxy is the same model as in Iodice et al. (2002a). We call $\mu$ the surface luminosity of the polar ring, obtained after the subtraction of the host galaxy model. The model for the ring is of the form:
\begin{equation}
\mu_{model}(x,z)=f(z) \exp \left( - \frac{(x-x_c(z))}{\Delta x} \right)
\end{equation}
where $z$ denotes the polar axis and $x$ denotes the axis perpendicular to the polar ring plane. 
Similarities between rings and rotating disks (Iodice 2001, 2002a) suggest that the polar ring may have an exponential profile perpendicular to its plane (as is the case on Fig.~\ref{snapmerg}). Yet, no reason implies that the polar ring radial profile $f(z)$ should be exponential, because the ring is at least partly depleted inside its radius $R$, $x_c(z)$ would be zero if the ring was exactly polar. It is not the case, and the ring is warped, too, so that we define for each $z$:
\begin{equation}
x_c(z)=\frac {\int x \mu(x,z) dx}{\int \mu(x,z) dx}
\end{equation}
Given that the exact radial profile of the ring can hardly be deduced from observations (the ring is seen edge-on), the most natural choice for $f(z)$ is simply:
\begin{equation}
f(z)=\mu(x_c(z))
\end{equation}
One free parameter, $\Delta x$, is then fitted to the HST image. 

This model does not study the ring radial distribution, but aims at exploring whether an exponential thickness of the ring accounts for the observed luminosity: if it is the case, matter observed far from the polar plane would belong to the faintest regions of the polar ring, if it is not the case, a third component (a stellar halo) would be formed by this matter.

\begin{figure}
	\centering
	\includegraphics[angle=270,width=8cm]{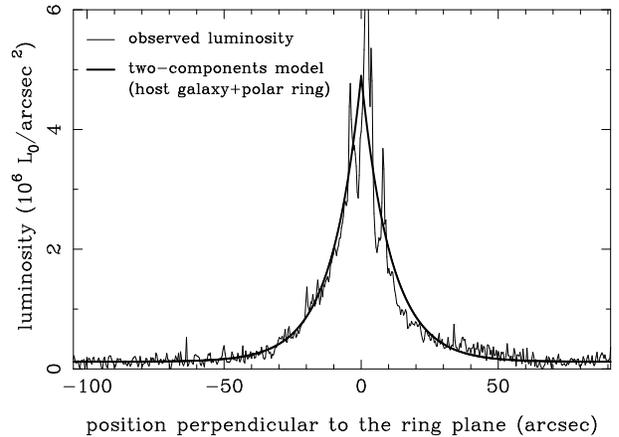}
	\caption{Profile of the polar ring perpendicular to its plane, observed at the ring radius $R$, and result of our two-component model: at this radius, the host luminosity is negligible, and the ring shows an exponential thickening. No stellar halo, that would be thicker than the ring, is detected.}
\label{coupe}
\end{figure}

The model fits observations well. For instance, at the ring radius $R$, an exponential thickening of the polar ring is really close to the observed profile (see Fig.~\ref{coupe}). No stellar halo then seems required, at least with the quality of our observations: no robust systematic deviation is observed between our model and the real luminosity. More precisely, if a stellar halo is present, it is below the noise: its surface luminosity is smaller than $\left| \mu(x,z)-\mu_{model}(x,z) \right|$. The relevant quantity is then:
\begin{equation}
\int_{\sqrt{x^2+z^2}<R} \left| \mu(x,z)-\mu_{model}(x,z) \right| \mathrm{d} x \mathrm{d}z
\end{equation}
It indicates the mass of the stellar halo inside the ring radius. As mentioned before, this mass is more than half of the stellar mass of the victim galaxy is nearly all our simulations. We then obtain that the stellar mass of the victim galaxy before the merging event should have been smaller than $3.6\cdot10^9$ M$_{\sun}$. 

The polar ring nowadays contains $8\cdot10^9$ M$_{\sun}$ of HI gas (Arnaboldi et al. 1997) and $4\cdot10^9$ M$_{\sun}$ of stars (Iodice et al. 2002a). The total mass of the polar ring is the gaseous mass of the victim galaxy: stars in the polar ring are formed after the ring formation. In fact, the gaseous mass of the victim galaxy may even be larger, if some gas is dispersed or converted into stars during the collision, before the ring formation. Then, the victim galaxy should have contained before the collision at the most $3.6\cdot10^9$ M$_{\sun}$ of stars and at least $12\cdot10^9$ M$_{\sun}$ of gas. It was thus characterized by:
\begin{itemize}
\item a gas mass fraction larger than 75\%
\item a visible mass larger than $15\cdot10^9$ M$_{\sun}$
\end{itemize} 
Such a gas mass fraction is found in both dwarf and low surface brightness (LSB) galaxies. As the total mass excludes dwarf galaxies, we conclude that if NGC~4650A is the result of a merging event, the victim galaxy should have been a LSB galaxy. 

This does not totally rule out the merging scenario. Yet, the accretion scenario was already most likely to occur. A supplementary constraint to the merging scenario is now that one of the invoked galaxies should have been an LSB. This constraint is not present in the accretion scenario, for no stars are accreted, and no stellar halo around the polar ring is expected. Thus, the probability that a merging event has resulted in NGC~4650A is strongly reduced, so that NGC~4650A is much likely to be the result of a tidal accretion event.

	\subsubsection{Polar disk or polar ring?}\label{diskvsring}

Iodice (2001), and Iodice et al. (2002a) have proposed that the polar structure of NGC~4650A is more likely a disk than a ring. Both shapes are not easy to discriminate, for the polar structure is seen edge-on. Iodice et al. (2002a) presented this as an argument in favor of the merging scenario. It is true that the radius of polar rings just after their formation is slightly larger in the merging scenario, according to N-body simulations. Yet, as explained by the radial model presented in Appendix~\ref{model}, the formation of the polar structure really results in a ring, not in a disk. Exceptionally, as noticed by Bekki (1998) and confirmed by run A4, very low velocity collisions could result in polar disks, provided that the relative velocity of colliding galaxies of about 30 or 40 km.s$^{-1}$: with such a strong supplementary constraint, and also the constraint on the victim stellar content found just below, the probability of the merging scenario really becomes marginal. Thus, the polar structure of NGC~4650A is bound either to be a ring and not a disk, or to be a polar disk resulting of a previous polar ring, while it is very unlikely that a polar disk has formed directly around this host galaxy.

Indeed, we have noticed in Sect.~\ref{morph_accr} (see also Fig.~\ref{polardisk}) that the accretion scenario forms rings with moderate radial extents, but that these rings may get a large radial extent, or even become disk-like structures, when they are disturbed after their formation. The fact that NGC~4650A may have a polar disk, or at least a very extended ring, is thus not an argument against one of the formation scenarios: both scenarios may result in ring with large radial extent, or even disks.

	\subsubsection{Surrounding galaxies}
NGC~4650A belongs to a galaxy group. We have selected ``surrounding galaxies'' of NGC~4650A from the LEDA database: such galaxies are assumed to be distant from NGC~4650A of less than 45 minutes, and have radial velocity larger than 2507 and lower than 3307 km.s$^{-1}$ (NGC~4650A radial velocity is 2907 km.s$^{-1}$). Foreground and background galaxies are rejected. For these 15 galaxies, we find a radial velocity dispersion of 200 km.s$^{-1}$. If we assume isotropic velocities in the group of galaxies, the mean relative velocity of two colliding galaxies in a head-on collision would be 400 km.s$^{-1}$, which makes a major merger resulting from a head-on collision very unlikely: two colliding galaxies are bound not to merge in this region of the Universe.

NGC~4650A and the spiral galaxy NGC~4650 are at the same radial distance (the radial velocity of NGC~4650 is 2909 km.s$^{-1}$). The real distance between both galaxies is only about 6 times the optical radius of NGC~4650. A tidal accretion event may then have form the polar ring from NGC~4650, and may easily explain why the ring is massive, since NGC~4650 is much more massive than NGC~4650A host galaxy. NGC~4650 is thus a candidate donor galaxy.

\subsection{NGC~660}\label{obs660}
The main characteristic of NGC~660 is that the ring extends to about 45 degrees from perpendicular (Whitmore et al. 1990, Combes et al. 1992). This polar ring is thus likely to be unstable, even if self-gravity may stabilize it if massive enough (Arnaboldi \& Sparke 1994). The second noticeable feature is the presence of gas in the host galaxy: the host gas content in typical of a late-type spiral: Van Driel et al. (1995) have found $M_\mathrm{HI}/L_\mathrm{B}^0=0.3 M_{\sun}/L_{\sun,\mathrm{B}}$ in the host disk, and detected a large mass of molecular gas in the disk.

According to numerical simulations of the merging scenario, no ring is inclined of more than 25 degrees from the polar axis. The ring of NGC~660 is maybe unstable, but has formed and is observed, which is not reproduced in the merging scenario. The fact is that in the merging scenario, very inclined rings are not observed to be unstable, but are observed not to form at all. The accretion scenario forms rings with larger inclinations. Inclined rings are rarely observed, for they are probably unstable and disappear in a few Gyrs (but the stability depends on the exact ring mass, and the distribution of dark matter). This selection effect makes most polar ring appear nearly polar, yet inclined rings exist, as in NGC~660, or also in ESO~235-G58 (Iodice 2001), which requires an accretion mechanism.

We have shown that when a polar ring forms in a galaxy merger, a significant fraction of gas falls inside the host galaxy during the ring formation. Yet, it forms a small, central structure, in which gas is quickly converted into stars, and that never replenishes the whole host disk with gas. Moreover, the intruder cannot be a gas-rich galaxy: even if the ring radius $R$ is larger than the radius of the intruder, polar gas would have large radial motion making its radius change and cross equatorial gas, which would prevent the ring from forming. Thus, either the gas of the intruder or the gas of the victim cannot produce a gas-rich host galaxy. In a tidal accretion event, the problem is solved: polar gas is directly placed around the ring radius $R$, so that it may never encounter the gas of the host galaxy, provided that the ring radius $R$ is larger than the host gas component radius, which is the case in NGC~660. Constraints on the gas content of the host galaxy are weaker in the accretion scenario than in the merging scenario, and the gas-rich host galaxy of NGC~660 may result only from a tidal accretion event. 

\begin{figure*}
	\centering
	\includegraphics[angle=270,width=7cm]{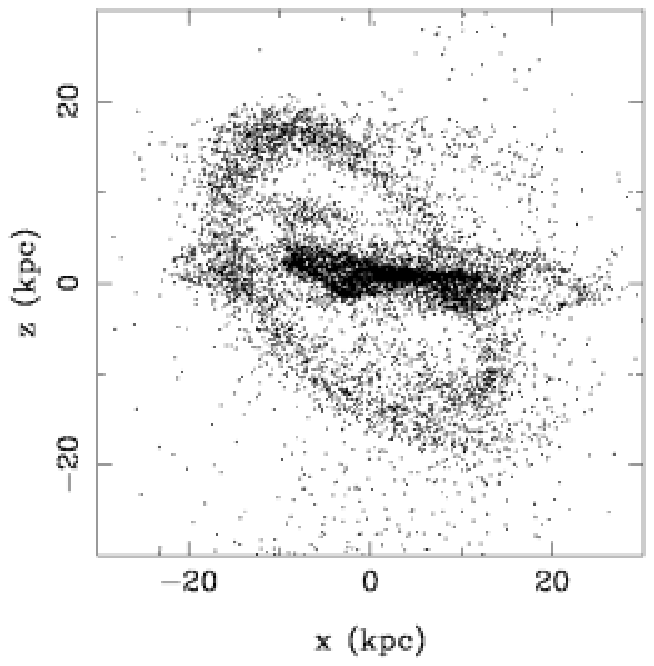}
	\hspace{1cm}
	\includegraphics[angle=270,width=7cm]{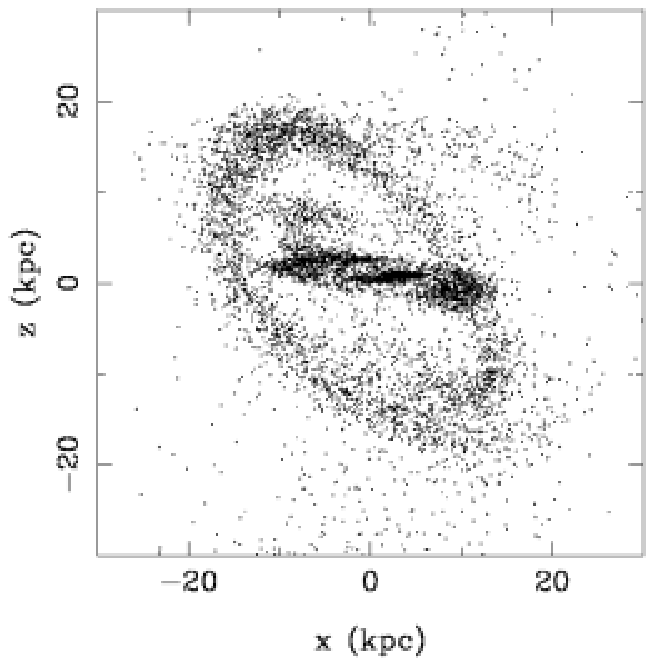}
	\caption{Run C26 : gas+stars (left) -- gas (right). The polar ring mass is 30\% of the host mass, the ring is 42 degrees from polar, and the host is gas-rich.}
\label{f660}
\end{figure*}

Finally, both characteristics of NGC~660 (inclined ring, gas-rich host galaxy) cannot be reproduced by our simulations of merging events. On the contrary, they are in agreement with the results of the accretion scenario. On Fig.~\ref{f660}, we show the result of a simulation of accretion, in which the polar ring is highly inclined (42 degrees) and the host galaxy is gas-rich (run C26). This numerical result is quite similar to NGC~600. The ring mass is here 30\% of the visible mass, and the polar ring is observed during 2.5 Gyrs in a nearly spherical dark halo: this ring is unstable but needs time to dissolve. Van Driel et al. (1995) have found that the ring mass in NGC~660 was even more important (40\% of the visible mass), so that the polar structure of NGC~660 will probably persist during some Gyrs after its formation.

As noticed by Iodice (2001), most polar rings are nearly polar. This could a priori be explained only by the merging scenario. Actually, this could also be explained by the accretion scenario; if the host potential is flat enough in most PRGs, inclined rings will be dissolved quickly, and will rarely be observed. So, the fact that polar rings are nearly polar does not argue for the merging scenario, at least until the shape of galactic potentials is known. Accretion may form many nearly polar rings, and some more inclined structures, rarely observed for they are less stable, and also as they are confused with barred galaxies (see Iodice 2001 for ESO~235-G58).

\subsection{NGC~2685}
NGC~2685, the Helix galaxy, has been studied in detail by Eskridge \& Pogge (1997). Observations of a large metallicity, color indices, and molecular gas were claimed to argue against the idea that the ring forms from tidally captured dwarf galaxies, and that the polar ring would be a long-lived self-gravitating structure instead. Yet, we have shown that the accretion scenario does not require the donor to be a dwarf galaxy, and that a massive donor can form a massive (thus self-gravitating) polar ring; moreover we also show that most polar rings formed by tidal accretion are long-lived. Thus, the results of Eskridge \& Pogge (1997) do not seem to argue against the accretion scenario, but only to require that the donor was rather massive and the ring is stable.

An important feature of NGC~2685 is the presence of an outer, equatorial ring. The host galaxy is not gas-depleted, as is the case for NGC~660, but contains gas at radii larger than the polar ring (Schinnerer \& Scoville 2002). 

We have already explained that the presence of large amount of gas in the host galaxy is not accounted for by the merging scenario. In the case of NGC~2685, the host galaxy is not really gas-rich, thus its gas content is not in contradiction with the merging scenario, provided that the victim disk has contained enough gas (see Sect.~\ref{constr_merg}). Yet, polar gas would sweep every radius before the polar ring forms (see Fig.~\ref{model33}, \ref{mot33}), which would compel equatorial gas to join the polar ring (as observed in run B13), in the same way as a larger quantity of equatorial gas would make the polar gas fall inside the equatorial plane. An outer equatorial ring and a polar ring -- more generally equatorial and polar gas -- cannot coexist after a merging event: one of the gaseous components (equatorial/polar) will join the other one, depending on their relative mass. Instead, two orthogonal rings may result from an accretion event, provided that they have not the same radius: the relevant aspect of this scenario is that polar gas is directly placed around the ring final radius $R$. Polar gas will then deplete the equatorial disk from its gas around $R$, but will not encounter gas at other radii. An outer ring may then be observed at radii larger than $R$, as well as gas at radii smaller than $R$ as is observed in NGC~660.

\section{Polar rings and dark matter}\label{s7}

\subsection{Dark halos of polar ring galaxies}
\subsubsection{Dark matter around donor galaxies}
The formation of a polar ring according to the accretion scenario is based on the development of tidal tails, which is influenced by the potential of the donor galaxy (Dubinski et al. 1996). As shown by Dubinski et al. (1999), some distributions of dark matter may inhibit the formation of tidal tails. This could prevent the accretion scenario from occurring, or make it much less frequent. However, tidal tails around interacting galaxies are frequently observed in the Universe. Thus, it seems that the results of Dubinski et al. (1996, 1999) should be interpreted in the following way: observations of tidal tails provides some constraints on dark halo shapes, and dark distributions that prevent tidal tails from forming are in fact rare, so that the accretion scenario actually occurs. 

\subsubsection{Dark matter around polar ring galaxies}
In our simulations of the merging scenario, when both galaxies are initially embedded in spheroidal halos, the final dark halo is spheroidal, and its flattening is most of the time smaller than E4. The dark-to-visible mass ratio is never larger than in preexisting spiral galaxies. 

According to the accretion scenario, when galaxies are initially embedded in spheroidal halos, the final halo is directly related to the preexisting halo of the host galaxy, slightly more spherical because of the tidal interaction. The dark-to-visible mass ratio in PRGs is again not larger than in spiral galaxies.

Iodice et al. (2003) have shown that polar ring velocities are abnormally large, when compared to the Tully-Fisher relation for spiral galaxies. The dark halo could then be either flattened along the polar ring, or spheroidal and very massive: the dark-to-visible mass ratio inside the optical radius would be as large as 3.5. Yet, velocities observed in host disk are not abnormally large (e.g. Combes \& Arnaboldi 1996), which led Iodice et al. (2003) to rule out spherical, very massive, halos. Our simulations confirm that the dark-to-visible mass ratio in PRGs cannot be generally larger than 1 (i.e. not larger than in spiral galaxies), which provides another argument against very massive dark halos. For this reason, only dark structures that are flattened towards polar rings account for observational results of Iodice et al. (2003), as already found by Combes \& Arnaboldi (1996) in the case of NGC~4650A. 

Dark halos flattened towards polar rings may be form during accretion 
events, provided that some dark matter is cold molecular gas (Pfenniger \& 
Combes 1994, Pfenniger et al. 1994). Under this hypothesis, the ideas that 
polar rings are issued of tidal accretion events and that dark halos of 
PRGs are flattened towards most polar rings are then compatible. On the 
other hand, when no dark matter is cold-gas, most halos are spherical or 
slightly flattened towards the host galaxies, and rarely toward the polar 
ring; which is not supported by observations (Iodice et al. 2003).

\section{Conclusion}\label{s8}
Our numerical simulations have confirmed that both scenarios are rather robust, and possibly result in realistic polar structures. The accretion scenario appears to be much more likely than the merging scenario: there are strong constraints for a merging event to result in a PRG. However, the merging scenario is robust enough to account for the fraction of PRG that have been detected until now.

Comparison of numerical results and observational properties tend to indicate that polar rings are actually the consequence of tidal accretion events. Identifying donor galaxies (when they have not merged with the host after ring formation) would be a confirmation, yet numerical simulations indicate that no obvious link remains between the donor and the ring, even only 1 or 2 Gyrs after the ring formation. Deep photometric studies, aiming at detecting the stellar remnant of the victim disk in the merging scenario, may bring another confirmation that PRGs are the result of tidal accretion events.

We also notice that the accretion scenario may form massive rings, as the ring of NGC~4650A. The donor is not required to be less massive than the future host, but could be more massive, and a large fraction of the donor gas could be accreted in the polar ring. Our simulations also suggest that most polar rings are long-lived structures, stabilized by self-gravity, independently of the formation scenario. Some inclined rings exist, as NGC~660. Such rings are likely to be unstable, but need time to be dissolved. Only tidal accretion can produce such inclined structures

We finally conclude that at least most PRGs -- and possibly all of them -- are the result of tidal accretion events. The two most recent studies of the distribution of dark matter in PRGs (Combes \& Arnaboldi 1996, Iodice et al. 2003), based on two different approaches, conclude that dark halos are flattened towards polar rings in most PRGs. This is fully compatible with the accretion scenario, and suggests that some dark matter is cold molecular gas.


\appendix

\section{Models for galaxies}\label{append}
\subsection{Disks}
 
We model stellar and gaseous disks with Toomre-Kuzmin disks, and a vertical distribution:
\begin{equation}
\rho(r,z)=\sigma(r) \mathrm{sech}^2\left(\frac{z}{h_0}\right)
\end{equation}
where $\sigma(r)$ is the surface density of the disk and $h_0=1$ kpc. The initialization of particles velocities for the disk is described in Bournaud \& Combes (2002).

\subsection{Dark halos, bulges, and elliptical galaxies}

Halos are first built as Plummer spheres and given an isotropic velocity dispersion corresponding to an isothermal virialized system. We sometime use flattened oblate structures of axis ratio $e$: the vertical position of each particle is then reduced by factor $e$ from the initial Plummer sphere. The anisotropy in the final system should then verify (Becquaert \& Combes 1997):
\begin{equation}\label{anis1}
e_v=\sqrt{ 1-{\frac{ 2(1-e^2)\left( \frac{1}{\sqrt{1-e^2}} -\arcsin(e)/e \right) }{\arcsin(e)/e-\sqrt{1-e^2}}} }
\end{equation}
where $e_v$ is defined by:
\begin{equation}\label{anisotr}
v_x^2=v_y^2=\frac{v_z^2}{\sqrt{1-e_v^2}}
\end{equation}
where $(x,y)$ is the disk plane, and $z$ the direction of the minor axis of the oblate dark halo. The kinetic energy is computed to keep the system virialized, velocities are then distributed to particles as if the system was isothermal, and oriented in space according to Eq.~\ref{anisotr}. The scale-length of dark halo is computed to insure a flat rotation curve in the outer disk.

Bulges are built in the same manner as dark halos. They are generally assumed to be E5 spheroidal systems.

Elliptical galaxies are first built as E0 systems according to the model of Jaffe (1983). They are then converted to oblate systems and virialized, exactly as was done for dark halos. Prolate systems have not been considered, for the shape of the host galaxy is not a crucial point in our study, and most host galaxies are disk galaxies and not ellipticals.

\begin{figure*}
	\sidecaption
	\includegraphics[width=6cm]{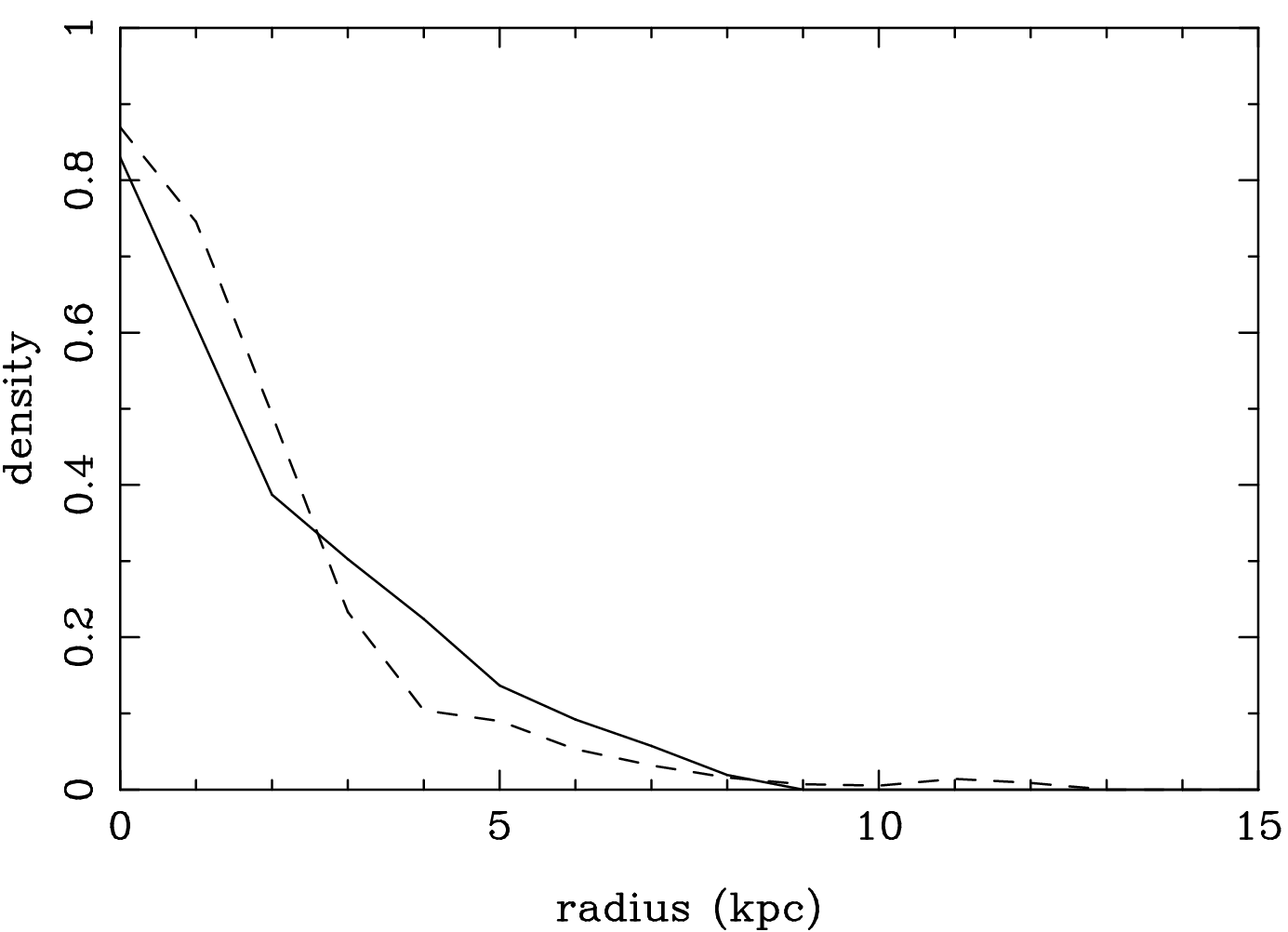}
	\includegraphics[width=6cm]{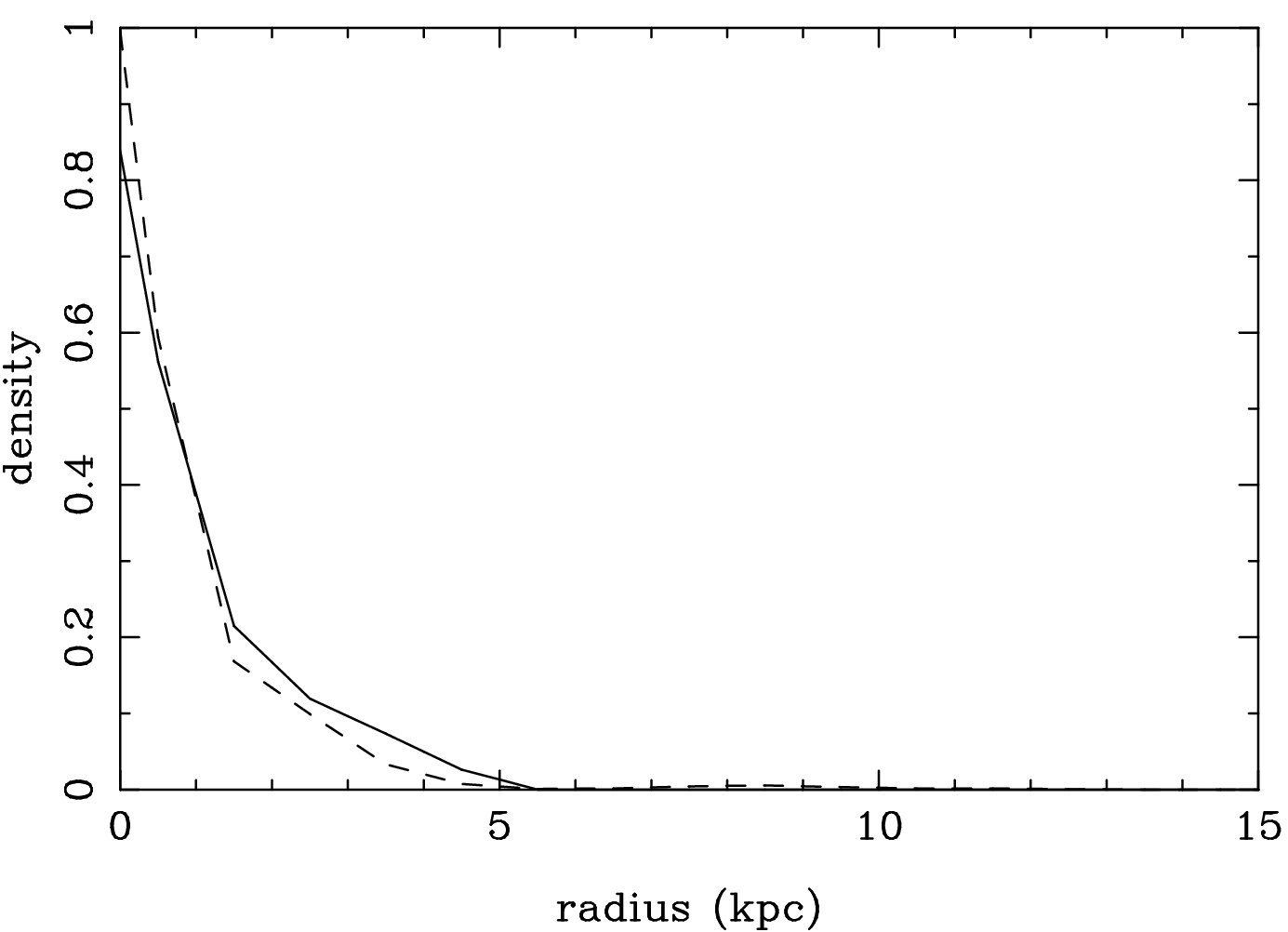}
	\caption{Relaxation of the model for elliptical galaxies: example of an E5 oblate system. The mass distribution along a major axis (left) and the minor axis (right) is shown at $t$=0 (solid line) and $t$=1 Gyr (dashed line). The system naturally evolves to a different state, but the change in its scale-lengths and axis ratio is not significant in our study.}
	\label{relax}
\end{figure*}

Eq.~\ref{anis1} is exact for infinite Plummer spheres. We use it for 
truncated distributions, and make the approximation that it is still 
correct for the model of Jaffe (1983). We use anisotropic velocity dispersions around the outer border of mass distributions, for particles with radial velocities would have escaped. Deviations from infinite Plummer spheres make the distribution of positions and velocities evolve to a relaxed state, that is close to the initial state. Even if we employ initial bulges, halos, and ellipticals, that are not stationary distribution, they have only slightly evolved when the galaxy interaction or mergers occurs. The mass distribution of an isolated elliptical galaxy is shown at $t$=0 and $t$=1 Gyr on Fig.~\ref{relax}.

\section{A model for the ring formation in the merging scenario}\label{model}

The reason for which a ring appears in this scenario is not obvious. The dissipative nature of the gas makes it settle in a plane, that is nearly polar for the initial angular momentum of the gas from the victim galaxy was polar, but does not simply explain why most of this gas tends to gather in a ring. On several runs, no resonance has been found between the potential of the host galaxy and the radius of the polar ring (this may have accounted for a ring formation). So, another explanation for the formation of the ring has to be found.

When polar collisions occur at large velocities, the two galaxies do not merge, and the victim disk becomes a cartwheel-like ring galaxy. Kinematical models succeed in explaining the formation of such extending rings that contain both stars and gas (Appleton \& Struck-Marcell 1996). The process is that the crossing of the intruder galaxy stimulates radial motions of the victim components. These radial motions occur at about the epicyclic frequency $\kappa(r_0)$ for a component initially at the radius $r_0$. All the components show simultaneous radial motions with coherent phases, which results in caustic waves, an makes an expanding ring appear. Such collisional transient rings are observed in our simulations. They do not appear to be directly related to the polar ring, for the polar ring is not a remnant of an expanding ring. However, the presence of caustic waves indicates that they may play a role in the formation of the polar ring. We modify the previous model, in order to account for the gas dissipative nature, and show that this enables us to explain why the gas gathers in a ring rather than a disk.

As studied in detail by Appleton \& Struck-Marcell (1996), we describe the galactic collision according to the impulse approximation, compute the epicyclic frequencies in the initial victim disk, and assume that the amplitude of radial motions varies as a power-law with radius. The model is unidimensional, and account only for the radius in the polar ring plane. This impulse approximation is questionable, for it consists to neglect the fact that the intruder may oscillate around the victim disk before merging. Yet, we have tested a model with two impulses, and the final results remained unchanged; moreover our physical interpretation clearly shows that what is important is to stimulate radial motions, while their exact amplitude is not crucial, as is the case when one studies the nature of collisional rings. Moreover, N-body simulations show that the ring radius is the same when the intruder oscillates around the victim disk several times.

Gas dissipation is added in this model. We assume that the dissipation rate is proportional to both the gas density and the gas radial velocity dispersion, and that the dissipation time-scale for the density and velocity dispersion of a galactic disk should be a few Gyrs. We then compute the mass density as a function of radius and  time. 

To achieve this model, we distribute particles on a radial axis, according to the initial profile of the victim disk. At $t=0$, radial sinusoidal motions are initialized, with a frequency $\kappa(r_0)$ for a particle initially at $r_0$, and an amplitude $A(r_0)$. $\kappa$ is computed in the potential of the intruder and the victim halo (since the victim disk is dispersed), and $A(r_0)$ is given by:

\begin{equation}
A(r_0)=A_0 \left( \frac{r_0}{a} \right)^{-m}
\end{equation}

According to Appleton \& Struck-Marcell (1996), values of $A_0$ and $m$ are not crucial for a qualitative discussion. We have generally chosen $A_0=0.6$ (for large radial amplitude are observed in simulations) and $m=0$ (i.e. a constant perturbation amplitude). 

At any instant $t$, a particle at radius $r$ undergoes a dissipative force:

\begin{equation}
F_\mathrm{d}=\frac{<v(r)>-v}{\tau}
\end{equation}
where $v$ is the radial velocity of the particle, $<v(r)>$ the mean radial velocity of particles at radius $r$, and $\tau=5$ Gyrs.

The effect is to change its velocity from $v$ to $v'$. The mean radius $r_0$ of the motion of the particle is then changed to $r_1$ according to:
\begin{equation}
r(t) = r_0+A(r_0)\cos\left( \kappa(r_0)t \right) = r_1+A(r_1)\cos\left( \kappa(r_1)t \right)
\end{equation}
\begin{equation}
v(t) = -A(r_0)\kappa(r_0)\sin\left( \kappa(r_0)t \right)
\end{equation}
\begin{equation}
v'(t) = -A(r_1)\kappa(r_1)\sin\left( \kappa(r_1)t \right)
\end{equation}
the phases of the particle motion is assumed not to change, and remain equal to 0, the mean radius $r_0$ is varied instead. The effect is that when a particle undergoes a dissipative event at radius $r(t)$, the mean radius of the particle motion, $r_1$, will be closer to $r(t)$ than $r_0$. Particles tend to gather where strong dissipation occurs.

\begin{figure}
	\centering
	\includegraphics[angle=270,width=8cm]{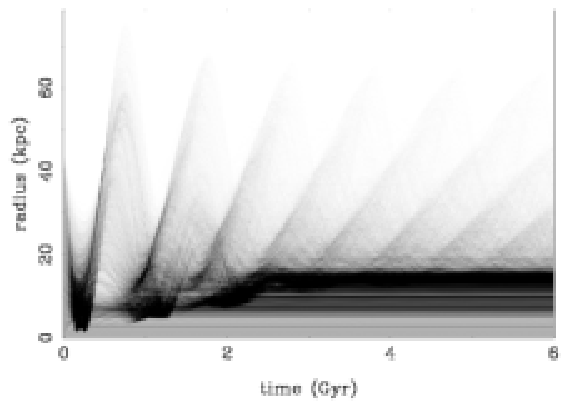}
	\caption{Radial model for the formation of a polar ring. The gas density is plotted in grey-scale as a function of time and radius. The physical parameters correspond to run A9, shown on Fig.~\ref{snapmerg}, in which the ring mean radius is $R=15$ kpc.}
\label{model33}
\end{figure}

The result for parameters corresponding to the run shown on Fig.~\ref{snapmerg} is displayed on Fig.~\ref{model33}: at short times, expanding caustic waves can be seen; a stationary ring appears later, and its radius is in agreement with what was expected from the N-body simulation.

\begin{figure}
	\centering
	\includegraphics[angle=270,width=8cm]{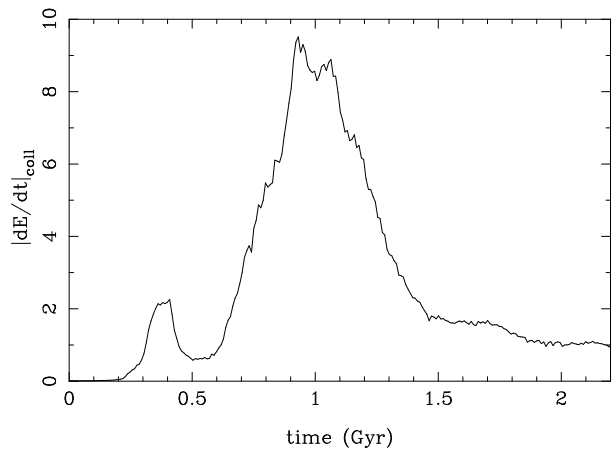}
	\caption{Energy lost by gas during collisions as a function of time, for the run shown on Fig.~\ref{snapmerg}, and related to run A9 (Fig.~\ref{model33}). As expected from the interpretation of our model, as strong dissipative event occurs before the accumulation of gas in a ring, around $t=1-1.2$ Gyrs.}
\label{ep33}
\end{figure}

\begin{figure}
	\centering
	\includegraphics[angle=270,width=8cm]{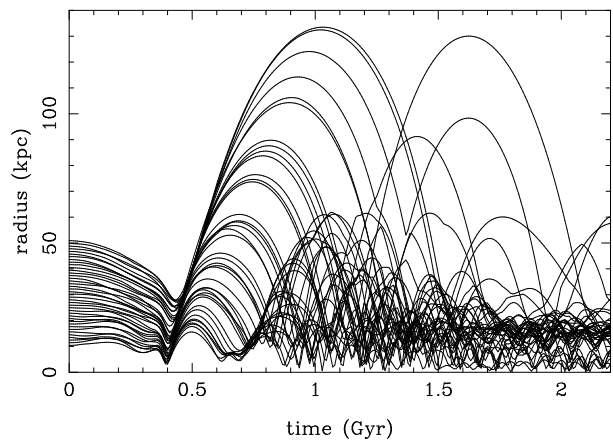}
	\caption{Particles radial motions in run A9: formation of a polar ring of mean radius $R=15$ kpc.}
\label{mot33}
\end{figure}

This axysimmetric model clearly confirms that the polar ring is not related to a resonance. However, it still provides no simple justification of the value of the ring radius. If we compute the mean angular momentum of the gas in the victim disk, and the polar rotation curve in the final system, we find the ring radius correspond to the mean initial angular momentum. This leads to a simple interpretation of the ring radius : radial motions stimulated by the intruder are first in phase and form transient collisional rings, but are later out of phase, which leads component to undergo strong dissipative events and to exchange their angular momentum, so that every component finally gets a momentum nearly equal to the initial mean momentum, which in terms of radius corresponds to the gathering of the gas around a given radius. So, radial motions enable exchanges of angular momentum that are responsible for the formation of the ring. A strong dissipative event should then occur before the ring formation, which is confirmed by numerical simulations (see Fig.~\ref{ep33}). This description explains why the ring radius depends largely on the gas radial extent in the victim galaxy. The formation of a polar ring from particles radial motions during the N-body simulation corresponding to the model on Fig.~\ref{model33} is shown on Fig.~\ref{mot33}

If $V(r)$ is the circular velocity as a function of radius in the victim's disk (before the collision) and $V'(r)$ is the mean circular velocity in the polar plane after the merging, the initial angular momentum per unit mass is:
\begin{equation}
L=\frac{\int_0^\infty r^2 V(r) \mu(r) dr}{\int_0^\infty r \mu(r) dr}
\end{equation}
where $\mu(r)$ is the gas surface density in the victim's disk (we here neglect velocity dispersion). The final mean radius of the polar ring, $R$, should then verify:
\begin{equation}
R V'(R)=L
\end{equation}
If we note $\delta R$ the difference between the radius predicted by our model and the radius measured in simulations, the mean value of $\delta R$/$R$ over the 34 studied simulations is 0.15. Given the good agreement of the radial model and N-body simulations, our interpretation of the ring formation seems robust: radial motions caused by the intruder and inelastic collisions enable gas clouds to exchange angular momentum, which makes them gather around a radius $R$. 

It seems important to notice that Iodice et al. (2002a) proposed that some polar structures should be regarded as polar disks rather than polar rings. On the other hand, it seems that the merging scenario in fact nearly always forms a ring distributed around a well determined radius $R$. This is discussed in the light of both formation scenarios in Sect.~\ref{diskvsring}.

\begin{acknowledgements}
We gratefully acknowledge Enrica Iodice and Magda Arnaboldi for 
providing us with HST data of NGC~4650A. This work has benefited from 
comments of an anonymous referee.
The computations in this work have been carried out on the Fujitsu
NEC-SX5 of the CNRS computing center, at IDRIS. 
We have made use of the LEDA database (http://leda.univ-lyon1.fr). 
This research has made use of the NASA/IPAC Extragalactic Database 
(NED) which is operated by the Jet Propulsion Laboratory, California 
Institute of Technology, under contract with the National Aeronautics 
and Space Administration. We have also made use of the Digitized Sky 
Survey. The Digitized Sky Survey was produced at the Space Telescope 
Science Institute under U.S. Government grant NAG W-2166. The images 
of these surveys are based on photographic data obtained using the 
Oschin Schmidt Telescope on Palomar Mountain and the UK Schmidt 
Telescope. The plates were processed into the present compressed 
digital form with the permission of these institutions.
\end{acknowledgements}


\begin{thebibliography}{}
\bibitem[1996]{Appleton96}Appleton, P. N., \& Struck-Marcell, C. 1996, Fundam. Cosmic Phys., 16, 111
\bibitem[1993]{Arnaboldi93}Arnaboldi, M., Capaccioli, M., Cappellaro, E., et al. 1993, A\&A, 267, 21
\bibitem[1994]{Arnaboldi94}Arnaboldi, M., Sparke, L. S. 1994, AJ, 107, 958
\bibitem[1997]{Arnaboldi97}Arnaboldi, M., Oosterloo, T., Combes, F. et al. 1997, AJ, 113, 585
\bibitem[1997]{Bekki97}Bekki, K. 1997, ApJ, 490L, 37
\bibitem[1998]{Bekki98}Bekki, K. 1998, ApJ, 499, 635
\bibitem[1997]{Becquaert97}Becquaert, J.-F., Combes, F. 1997, A\&A, 325, 41
\bibitem[2002]{Bournaud02}Bournaud, F., Combes, F. 2002, A\&A, 392, 83
\bibitem[1997]{Broeil97}Broeils, A. .H, \& Rhee, M.-H. 1997, A\&A, 324, 877
\bibitem[1985]{Combes85}Combes, F., Gerin, M. 1985, A\&A, 150, 327
\bibitem[1992]{Combes92}Combes, F., Braine, J., Casoli, F., et al. 1992, A\&A, 259L, 65
\bibitem[1996]{Combes96}Combes, F., Arnaboldi, M. 1996, A\&A, 305, 763
\bibitem[1994]{Curir94}Curir, A., Diaferio, A. 1994, A\&A, 285, 389
\bibitem[1974]{Danziger74}Danziger, I. J., \& Schuster, H. E. 1974, A\&A, 34, 301
\bibitem[1996]{Dubinski96}Dubinski, J., Mihos, J. C., Hernquist, L. 1996, ApJ, 462, 576
\bibitem[1999]{Dubinski99}Dubinski, J., Mihos, J. C., Hernquist, L. 1999, ApJ, 526, 607
\bibitem[1997]{Eskridge97}Eskridge, P., Pogge, R. W. 1997, ApJ, 486, 259
\bibitem[2002]{Gallagher02}Gallagher, J. S., Sparke, L. S., Matthews, L. D., et al. 2002, ApJ, 568, 199
\bibitem[1998]{Haud88}Haud, U. 1988, A\&A, 198, 125
\bibitem[1997]{Horellou97}Horellou, C., Booth, R. 1997, A\&AS, 126, 3
\bibitem[2001]{Horellou01}Horellou, C., Combes, F. 2001, Ap\&SS, 276, 1141
\bibitem[1983]{Jaffe83}Jaffe, W. 1983, MNRAS, 202, 995
\bibitem[2001]{Iodice01}Iodice, E. 2001, Ph.D.~Thesis, SISSA, Trieste
\bibitem[2002]{Iodice02A}Iodice, E., Arnaboldi, M., De Lucia, G., et al. 2002a, AJ, 123, 195
\bibitem[2002]{Iodice02B}Iodice, E., Arnaboldi, M., Sparke, L. S., Freeman, K. C. 2002b, A\&A, 391, 117
\bibitem[2002]{Iodice02C}Iodice, E., Arnaboldi, M., Sparke, L. S., Freeman, K. C. 2002c, A\&A, 391, 114
\bibitem[2002]{Iodice03}Iodice, E., Arnaboldi, M., Bournaud, F., et al. 2003, ApJ, in press (astro-ph/0211281)
\bibitem[1977]{James77}James, R.A. 1977, J. Comput. Phys., 25, 71
\bibitem[1994]{Pfenniger94a}Pfenniger, D., Combes, F., Martinet, L. 1994, A\&A, 285, 79
\bibitem[1994]{Pfenniger94b}Pfenniger, D., Combes, F. 1994, A\&A, 285, 94
\bibitem[1997]{Reshetnikov97}Reshetnikov, V., Sotnikova, N. 1997, A\&A, 325, 933
\bibitem[1994]{Richter94}Richter, O.-G., Sackett, P. D., Sparke, L. S. 1994, AJ, 107, 99
\bibitem[1983]{Sancisi83}Sancisi, R. 1983 Internal Kinematics and Dynamics of 
Galaxies, IAU Symposium, 100, 55-62
\bibitem[1990]{Sackett90}Sackett, P. D., Sparke, L. S. 1990, ApJ, 361, 408
\bibitem[1994]{Sackett94}Sackett, P. D., Rix, H.-W., Jarvis, B. J., Freeman, K. C. 1994, ApJ,436, 629
\bibitem[1981]{Schwarz81}Schwarz, M. P. 1981, ApJ, 247, 77
\bibitem[2002]{Schinnerer02}Schinnerer, E., Scoville, N. Z. 2002, ApJ, 577, L103
\bibitem[1983]{Schweizer83}Schweizer, F., Whitmore, B. C., Rubin, V. C. 1983, AJ, 88, 909
\bibitem[1986]{Sparke86}Sparke, L. S. 1986, MNRAS, 219, 657
\bibitem[1995]{vanDriel95}van Driel, W., Combes, F., Casoli, F., et al. 1995, AJ, 109, 942
\bibitem[2000]{vanDriel00}van Driel, W., Arnaboldi, M., Combes, F., Sparke, L. S. 2000, A\&AS, 141, 385
\bibitem[2002]{vanDriel02}van Driel, W., Combes, F., Arnaboldi, M., Sparke, L. S. 2002, A\&A, 186, 140
\bibitem[1987]{Whitmore87}Whitmore, B. C., McElroy, D. B., Schweizer, F. 1987, ApJ, 314, 439
\bibitem[1990]{Whitmore90}Whitmore, B. C., Lucas, R. A. 1990, AJ, 100, 1489
\end{thebibliography}
\end{document}